\documentclass[%
reprint,
superscriptaddress,
amsmath,amssymb,
longbibliography,
aip,jcp]{revtex4-2}

\usepackage{graphicx}
\usepackage{verbatim}
\usepackage[version=4]{mhchem}
\usepackage{xcolor}
\usepackage{physics}
\usepackage[normalem]{ulem}
\usepackage{pdfpages}
\usepackage{pgffor}
\usepackage{soul}
\usepackage{hyperref}
\usepackage{booktabs}
\usepackage{pdfpages}
\usepackage{pgffor}

\usepackage{bm}
\newcommand{\ocg}{\,\widetilde{\otimes}\,}
\newtheorem{prop}{Proposition}
\newtheorem{corollary}[prop]{Corollary}
\newcommand{\br}{{\bm r}}
\newcommand{\bq}{{\bm q}}
\newcommand{\mS}{S}

\makeatletter
\AtBeginDocument{\let\LS@rot\@undefined}
\makeatother

\begin{document}

\title{Representing spherical tensors with scalar-based machine-learning models}

\author{M. Domina}
\affiliation{Laboratory of Computational Science and Modeling, Institut des Mat\'eriaux, \'Ecole Polytechnique F\'ed\'erale de Lausanne, 1015 Lausanne, Switzerland}

\author{F. Bigi}
\affiliation{Laboratory of Computational Science and Modeling, Institut des Mat\'eriaux, \'Ecole Polytechnique F\'ed\'erale de Lausanne, 1015 Lausanne, Switzerland}

\author{P. Pegolo}
\affiliation{Laboratory of Computational Science and Modeling, Institut des Mat\'eriaux, \'Ecole Polytechnique F\'ed\'erale de Lausanne, 1015 Lausanne, Switzerland}

\author{M. Ceriotti}
\affiliation{Laboratory of Computational Science and Modeling, Institut des Mat\'eriaux, \'Ecole Polytechnique F\'ed\'erale de Lausanne, 1015 Lausanne, Switzerland}

\date{\today}

\begin{abstract}
Rotational symmetry plays a central role in physics, providing an elegant framework to describe how the properties of 3D objects -- from atoms to the macroscopic scale -- transform under the action of rigid rotations. Equivariant models of 3D point clouds are able to approximate structure-property relations in a way that is fully consistent with the structure of the rotation group, by combining intermediate representations that are themselves spherical tensors. The symmetry constraints however make this approach computationally demanding and cumbersome to implement, which motivates increasingly popular unconstrained architectures that learn approximate symmetries as part of the training process. 
In this work, we explore a third route to tackle this learning problem, where equivariant functions are expressed as the product of a scalar function of the point cloud coordinates and a small basis of tensors with the appropriate symmetry. 
We also propose approximations of the general expressions that, while lacking universal approximation properties, are fast, simple to implement, and accurate in practical settings.
\end{abstract}

\maketitle

\section{Introduction}
Symmetries underpin the laws that describe the universe at both macroscopic and microscopic scales, and have therefore traditionally been core constraints when developing physical and data-driven models.
Equivariance, a mathematical formalization of symmetry, ensures that physical laws and equations remain consistent under transformations such as rotations and translations. 
In particular, the equivariant fitting of microscopic observables has risen in popularity in the last decade, allowing for example to dramatically speed up workflows which would otherwise require expensive quantum electronic structure calculations.
When fitting quantum mechanical observables with machine learning models, the importance of rotational symmetry becomes obvious by observing that all the targets of interest are tensors of an appropriate rank, such as forces, dipoles, multipoles, stress tensors, polarizabilities, hyperpolarizabilities, just to mention a few. 
Since it is well known that a tensor can be decomposed into its irreducible representations (which all have a well-determined equivariant character) with respect to their behavior under rotation or inversion (elements of the group), then any tensorial quantity is intrinsically equivariant.
The decomposition into irreducible components is carried out by making use of the tools established in the theory of the coupling of angular momentum. 
In particular, the Clebsch-Gordan (CG) coefficients can be used to obtain all the irreducible components by a simple contraction, in analogy to the construction of multipolar spherical harmonics from the standard ones~\cite{VMK}. 
Crucially, the CG contraction is at the very core of most Machine-Learning (ML) architectures in this domain \cite{bart+10prl,bart+13prb,thom+15jcp,gris+18prl,ande+19nips,unke-meuw19jctc,will+19jcp,drau19prb,niga+20jcp,lewi+21jctc,musi+21cr,bazt+22ncomm,frank+22nips,bata+22nips,domi+22prb,NguyenLunghi2022,musa+23ncomm,bigi+24jcp,bochkarev+24prx}.

We base our discussion on the results of Ref.~\citenum{NEURIPS2021_f1b07759}, where it is proven that only scalar functions are required to describe a vectorial property in arbitrary dimension. 
In this work we will focus exclusively on $\mathbb{R}^3$ space, as we are mostly interested in describing point clouds in real space. The main result of Ref.~\citenum{NEURIPS2021_f1b07759} is that, given an equivariant vectorial function $\bm h$ depending on a set of input vectors $\{\br_i\}_{i=1}^n$, it is possible to find $n$ scalar functions $f_i$ depending on the same set of inputs (in particular, depending on products of powers of inner products of the input vectors, $(\br_i\cdot \br_j)$, as implied by the fundamental theorem of invariant theory~\cite{NEURIPS2021_f1b07759,shap16mms,weyl1946classical}), such that
\begin{equation}\label{eq:scalars_are_universal}
    \bm h(\{\br_i\}) = \sum_{j=1}^n f_j(\{\br_i\}) \br_j,
\end{equation}
namely this function is fully defined on the space spanned by its inputs. This property is a direct consequence of the equivariance of the function $\bm h$, which can be stated as follows: Given an element $Q$ of the orthogonal group $Q\in O(3)$ and a suitable representation of it, $\bm Q$, then it holds that 
    $\bm h(\{\bm Q\br_i\}) = \bm Q \bm h(\{\br_i\})$. 
However, when dealing with higher rank tensors, the results of Eq.~\eqref{eq:scalars_are_universal} are not easily generalizable.  This can be seen by means of the following (proper) rank 2 tensor
\begin{equation}\label{eq:scalars_are_universal_counter_example}
    \bm T(\br_1, \br_2) = (\br_1\times \br_2)\otimes (\br_1\times \br_2),
\end{equation}
where $\times$ and $\otimes$ are the usual cross-product and external product respectively. Clearly, this tensor lies on a space orthogonal to any combination $\br_i\otimes\br_j$, with $i,j=1,2$, while being equivariant with respect to the operator $\bm Q\otimes \bm Q$. 
As discussed in Ref.~\citenum{gregory2024learningequivarianttensorfunctions}, a generalization of \eqref{eq:scalars_are_universal} in terms of external products require the introduction of ``Kronecker-delta tensors''. 
While this result leads to an understanding of the relation between the input space of the tensor and its equivariance with respect to the action of a group, it does not address the case in which the tensor is given in terms of irreducible representations. 
In this work, we mainly focus on the irreducible representation of the groups $O(3)$ and $SO(3)$ with vectorial inputs, and we prove that a much simplified generalization can be obtained.
In particular, from expressions in terms of external products only, the irreducible representation is obtained by considering all possible Clebsch-Gordan (CG) contractions~\cite{VMK} of the vectors in the input.
Here, we prove that this is not necessary: 
not only are no additional tensorial terms present, such as the Kronecker-delta tensors, but also only the maximal CG contractions are needed. This completely removes the need to account for coupling schemes, which are among the most computationally demanding operations of a ML architecture.
Since, in most cases, irreducible representations are the object of study as they mirror the symmetries of the system, our results have the potential to introduce a new perspective for all methods that explicitly targets those representations.

An important point to mention is the results of Ref.~\citenum{NEURIPS2021_f1b07759} when the vectorial function is permutationally invariant under the swap of any of the vectors of the input. 
In this case Eq.~\eqref{eq:scalars_are_universal} takes the simplified form
\begin{equation}\label{eq:permut_inv_scalars_universal}
    \bm h (\{\br_i\}) = \sum_{j=1}^n f\big(\br_j, [\br_k]_{k\neq j}\big) \br_j,
\end{equation}
where now there is only one scalar function $f$ which is permutationally invariant with respect to all the arguments but the first one (the permutational invariance of the arguments is here indicated with square brackets). 
A similar simplification also applies to our results, with a dramatic reduction in the number of scalar functions that have to be considered in the expansion.
However, we will show that exploiting permutational invariance is still not enough to obtain a framework with favorable scaling with respect to the number of input vectors, especially when the irreducible representation is of high angular momentum. 

Thus, the main goals of this work are two-fold.
On the one hand we aim to show that results analogous to Eq.~\eqref{eq:scalars_are_universal} can be obtained for an equivariant representation written in the language of spherical harmonics and CG contractions. 
On the other, because of the impracticality of the theoretical results, we aim to provide useful approximations and a lightweight architecture that can target equivariant objects.
The manuscript is structured as follows. In \autoref{sec:methods} we give our main theoretical results, showing how previous results of Ref.~\citenum{NEURIPS2021_f1b07759} can be directly generalized to the case of spherical tensors, by means of the maximal couplings introduced in Eq.~\eqref{eq:maximal_coupling}. At the end of the same section we discuss more general consequences of the results, and the differences with the previous generalization given by Ref.~\citenum{gregory2024learningequivarianttensorfunctions}. As the theoretical results are non-practical for actual implementations, in \autoref{sec:practical} we investigate approximations that can be efficient, with minimal compromises on the generality of the problem. Finally, in \autoref{sec:results}, we report our tests on several case studies, to investigate the accuracy of the approximations and extensions of the methods.

\section{Methods}\label{sec:methods}

\subsection{Definitions}
We begin by presenting a few remarks and definitions. Given a general Cartesian tensor $\bm A$, we will indicate with $\bm A^L\in \mathbb{R}^{2L+1}$ its harmonic components belonging to one of its $(2L+1)$-dimensional irreducible decompositions. 
In general, there are several independent ways to define the irreducible components of a tensor depending on the chosen coupling scheme. 
However, any such distinction will be unessential for our treatment, and thus any further label required to uniquely identify the irreducible space of interest will be implied. 
In order to keep our treatment as readable as possible, we will define the CG contraction between the harmonic components of two tensors $\bm A$ and $\bm B$ as
\begin{equation}
    (\bm A^{L_1} \ocg \bm B^{L_2})_{\lambda\mu} := \sum_{M_1M_2} C^{\lambda\mu}_{L_1M_1L_2M_2} (\bm A^{L_1} \otimes \bm B^{L_2})_{M_1M_2}, 
\end{equation}
written in terms of the standard external product $\otimes$, where the integer indexes $M_1$ and $M_2$ satisfy the constraints $\abs{M_1}\leq L_1$ and $\abs{M_2}\leq L_2$. 
We will use throughout real spherical harmonics, and therefore the CG coefficients are those that are suitable to enact coupling with this convention. 
The CG contraction is either commutative or anti-commutative, following the rule
\begin{equation}
    (\bm B^{L_2} \ocg \bm A^{L_1})_{\lambda\mu} = (-1)^{L_1+L_2-\lambda}(\bm A^{L_1} \ocg \bm B^{L_2})_{\lambda\mu},
\end{equation}
inherited by the symmetries of the CG coefficients~\cite{VMK}. The expressions above also display the bilinearity of this contraction, namely
\begin{equation}
\begin{split}
        &\big((\alpha \bm A^{L_1} + \beta \bm B^{L_1}) \ocg \bm C^{L_2}\big)_{\lambda\mu} \\
        &= \alpha (\bm A^{L_1}\ocg \bm C^{L_2})_{\lambda\mu}+ \beta (\bm B^{L_1}\ocg \bm C^{L_2})_{\lambda\mu},
\end{split}
\end{equation}
for any real number $\alpha, \beta\in \mathbb{R}$. In the following we will focus mostly on 3-dimensional vectors: in terms of a harmonic representation vectors transforms as $\lambda = 1$ components. 
Using a common convention for real spherical harmonics, if we have a vector $\bm a \in \mathbb{R}^3$ with components $\bm a = (a_x, a_y, a_z)$, we can make the identifications $a_z = a_0$, $a_y = a_1$ and $a_x = a_{-1}$, which is consistent with the application of CG products for real-valued spherical harmonics. 
The CG contraction between two vectors $\bm a$ and $\bm b$ can be easily interpreted for $\lambda = 0$ and $\lambda =1$. 
\begin{figure*}[t!]
    \centering
    \includegraphics[width=\textwidth]{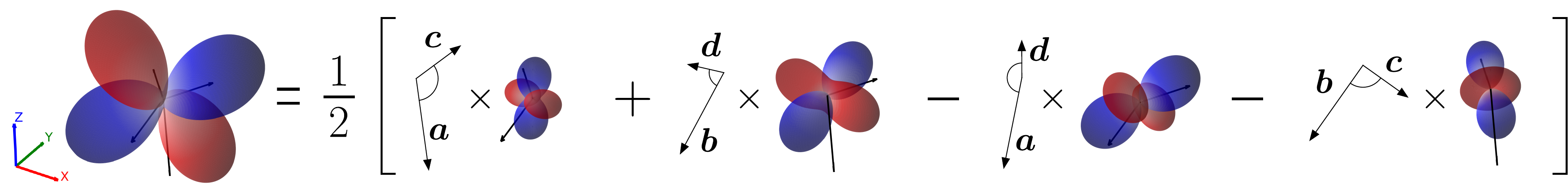}
    \caption{
    Graphical representation of Eq.~\eqref{eq:fundamental_identity}, with vectors $ \bm a = (2, -3, -1)$, $\bm b = (-2, 1, -3)$, $\bm c = (1, 2, 0)$, $\bm d = (-1, 1, 0.5)$. 
    The polar plots are obtained by evaluating the function 
    $f_{\bm T}(\bm{\hat{r}}) := \bm{\hat{r}}^T \bm T \bm{\hat{r}}$, where $\bm{\hat{r}}\in S^2$ is a versor of the unitary sphere, and $\bm T$ is the Cartesian representation of the $\lambda = 2$ tensors appearing in the equation. 
    The right-hand side is magnified by a factor $\sqrt{2}$ for visual purposes.}
    \label{fig:vector_relation}
\end{figure*}
Indeed, the contraction to the scalar ($\lambda=0$) space between two arbitrary vectors $\bm a$ and $\bm b$ is proportional to the scalar products between the vectors, as shown by the formula $(\bm a \ocg \bm b)_{00} = -(\bm a \cdot \bm b)/\sqrt{3}$. 
Instead, the Cartesian representation of the contraction to the $\lambda =1$ (pseudo-vectorial) space is proportional to the cross product, namely $(\bm a \ocg \bm b)_{1}=  -\mathrm{i} (\bm a \times \bm b)/\sqrt{2}$. 
The presence of the imaginary unit $\mathrm{i}$ is due to the fact that in the real spherical harmonic representations, the CG coefficients $C^{1\mu}_{1m_11m_2}$ are purely imaginary. 
Note also how the same relations holds also for the components of the standard spherical harmonic representation, as shown in Ref.~\citenum{Sakurai_Napolitano_2020}(Ch. 3.11).
Moreover, the contraction to the $\lambda=2$ space can also be easily obtained in its Cartesian form: it is sufficient to extract the traceless and symmetric part of the matrix obtained from the external product $\bm a \otimes \bm b$. These intuitive relations are crucial when investigating tensors from vectorial inputs and we will heavily use them in the next paragraph.

We now state an identity for the contraction of four arbitrary vectors $\bm a, \bm b, \bm c$ and $\bm d$ (see Ref.~\citenum{VMK}(Ch.~3) for similar identities):

\begin{widetext}
\begin{equation}\label{eq:fundamental_identity}
\begin{split}
\left((\bm a \ocg \bm b)_1\ocg(\bm c \ocg \bm d)_1\right)_{2}  = 
 \dfrac{1}{2}\Big[(\bm a \cdot\bm c)\big(\bm b\ocg\bm d\big)_{2}+(\bm b \cdot \bm d)\big(\bm a \ocg\bm c\big)_{2}
    - (\bm a \cdot \bm d)\big(\bm b\ocg \bm c\big)_{2}-(\bm b\cdot \bm c)\big(\bm a \ocg \bm d\big)_{2}\Big],
\end{split}
\end{equation}
\end{widetext}

This formula can be seen as a generalization of the well-known identity $ (\bm a \times \bm b)\cdot (\bm c\times \bm d) = (\bm a \cdot \bm c)(\bm b \cdot \bm d) - (\bm a \cdot \bm d)(\bm b \cdot \bm c)$ and, despite its simplicity, is of fundamental importance for this work. 
In words, it shows that a CG contraction to the space at $\lambda = 2$ of four vectors can be decomposed in scalar terms multiplied by pair-wise CG contractions, which are the simplest covariant objects that can be constructed from two vectors. 
This points in the direction of a generalization of the results presented in Ref.~\citenum{NEURIPS2021_f1b07759} (see Eq.~\eqref{eq:scalars_are_universal}), where the vectorial nature of the target was separated into vectors (the simplest equivariant object with $\lambda=1$), modulated by scalar contributions.
The proof of Eq.~\eqref{eq:fundamental_identity}, obtained by exploiting the theory of re-coupling of angular momenta, is shown in SI.

Given the identity above, we can build some intuition on how the example of Eq.~\eqref{eq:scalars_are_universal_counter_example} can be generalized to be in line with the results of Ref.~\citenum{NEURIPS2021_f1b07759}. Indeed, its projection onto the $\lambda = 2$ space can be written as
\begin{equation}\label{eq:example_projection_orthogonal}
\begin{split}
    \bm T_{2} (\br_1, \br_2)\, &\propto \, ((\br_1 \ocg \br_2)_1\ocg (\br_1 \ocg \br_2)_1)_{2} 
    \\
    &= \dfrac{1}{2}\Big(r_1^2 (\br_2 \ocg \br_2)_2 + r_2^2 (\br_1 \ocg \br_1)_2\Big) 
    \\
    &\quad - (\br_1 \cdot \br_2) (\br_1 \ocg \br_2)_{2},
\end{split}
\end{equation}
where the proportionality is realized by unessential constants and where we defined the length of the vectors as $r_i := \abs {\br_i}$. 
This shows how this tensor lies in the space spanned by the contractions $(\br_i\ocg \br_j)_2$ only. In this work we will show how this fact holds in general for any arbitrary angular momentum $\lambda$.

\subsection{Maximal coupling}
Here we shortly discuss the definition of the maximal coupling, which is at the core of our results. It is defined as the contraction of harmonic components to the highest allowed angular momentum. For example, in the case of $\lambda$ vectors, $\{\bm a_i\}_{i=1}^\lambda$, the maximal coupling is
\begin{equation}
    \label{eq:maximal_coupling}
    \begin{split}
    &(\overbrace{\bm a_1 \ocg \ldots \ocg \bm a_\lambda}^{\lambda\text{ times}})_{\lambda\mu}
    \\
    &:= \Big(\big(\cdots\big((\bm a_1 \ocg \bm a_2)_2 \ocg \bm a_3\big)_3\ocg \cdots \ocg \bm a_{\lambda-1}\big)_{\lambda-1}\ocg \bm a_\lambda\Big)_{\lambda\mu} 
    \phantom{:}
    \\
    &\phantom{:}= \sum_{m_1\ldots m_\lambda} \mathcal{C}^{\lambda\mu}_{m_1\ldots m_\lambda}\big((\bm a_1)_{m_1}\cdot\ldots\cdot (\bm a_\lambda)_{m_\lambda}\big),
    \end{split}
\end{equation}
obtained by contracting the vectors to the \textbf{maximum} $\lambda$ allowed by the CG contraction. Here, $\mathcal{C}^{\lambda\mu}_{m_1\ldots m_\lambda}$ is a shorthand for the components of the tensor $\mathcal{C}$ obtained by the contraction of the CG coefficients involved in the above definition, with $m_i = 0,\,\pm 1$. Since every vector contributes with a harmonic representation of degree $1$, the degree of the highest representation coincides with the number of vectors. The most important property of the maximal coupling is that it is commutative in the order of the vectors $\bm a_i$, which is a direct consequence of the fact that the highest coupling of two angular momenta is symmetric. In other words, the tensor $\mathcal{C}$ is totally symmetric under any swap of the $m_i$ indexes. 
For this reason, it is not necessary to specify any intermediate CG contraction and, crucially, \emph{any choice of coupling will produce the same result}.  
Finally, a useful shorthand is provided by the definition
\begin{equation}\label{eq:self-contraction}
    \bm a^{\ocg \lambda} := \underbrace{(\bm a \ocg\ldots \ocg \bm a)_{\lambda}}_{\lambda\text{ times}},
\end{equation}
namely the maximal CG coupling of the vector $\bm a$ with itself.
The maximal coupling will be the only kind of coupling that will be used in the rest of this work when contracting more than two vectors.

\subsection{Results}

The most important characteristic of an equivariant harmonic tensor $\bm T_{\lambda}$ of angular momentum $\lambda$, are transformations with respect to a rotation and with respect to inversion. The former is expressed by 
\begin{equation}\label{eq:rotation}
    R\, :\, T_{\lambda\mu}\longrightarrow \sum_{\mu'} D^{\lambda}_{\mu\mu'}(\mathcal{R})\,T_{\lambda\mu'},
\end{equation}
where $R$ is a rotation, $\mathcal{R}$ is its parametrization in terms of Euler angles and $\bm{D}^\lambda$ are the (real) Wigner $D$-matrices. Here $\mu$ and $\mu'$ takes integer values from $-\lambda$ to $\lambda$. Regarding the inversion (parity operation) we have two different behaviors for proper tensors and pseudotensors, namely
\begin{equation}\label{eq:parity}
\begin{split}
    P\,:\, T_{\lambda\mu} \longrightarrow (-1)^\lambda T_{\lambda\mu}, &\qquad\text{for proper tensors,}\\
    P\,:\, T_{\lambda\mu} \longrightarrow -(-1)^\lambda T_{\lambda\mu},&\qquad\text{for pseudotensors.}
\end{split}
\end{equation}
For example, a scalar ($\lambda =0$) is invariant and a proper vector ($\lambda = 1$) changes sign, while a pseudoscalar (like the determinant obtained from three vectors) changes sign and a pseudovector (like the cross product of two proper vectors) stays unchanged.
Given the different behavior under inversion, our investigation treats the proper and pseudo cases separately.

The core objective of any symmetry-consistent approximation, as investigated in this work, is to achieve equivariance of the tensor with respect to transformation of its inputs. 
Explicitly, let us consider an element $Q\in O(3)$ of the orthogonal group (where $Q$ can describe a rotation, and inversion, or a combination of both) and its representation on the space of angular momentum $\lambda$, here indicated with $\bm Q^{(\lambda)}$. Then a harmonic tensor $\bm T_\lambda$, of inputs $\{\br_i\}$ is said to be equivariant if it holds that
\begin{equation}\label{eq:equivariance}
    \bm Q^{(\lambda)}\bm T_{\lambda}(\{\br_i\}) = \bm T_{\lambda}(\{\bm Q^{(1)}\br_i\}).
\end{equation}
We remark that the the explicit form of $\bm Q^{(\lambda)}$ depends on the space on which we are acting upon, as shown in Eqs.~\eqref{eq:rotation} and \eqref{eq:parity}. 
Indeed, while a rotation is always represented by the real Wigner $D$-matrices, an inversion is represented by means of the first line or the second line of Eq.~\eqref{eq:parity}, if $\bm T_\lambda$ is a proper tensor or a pseudotensor, respectively.

\begin{figure*}[t!]
    \centering
    \includegraphics[width=.92\textwidth]{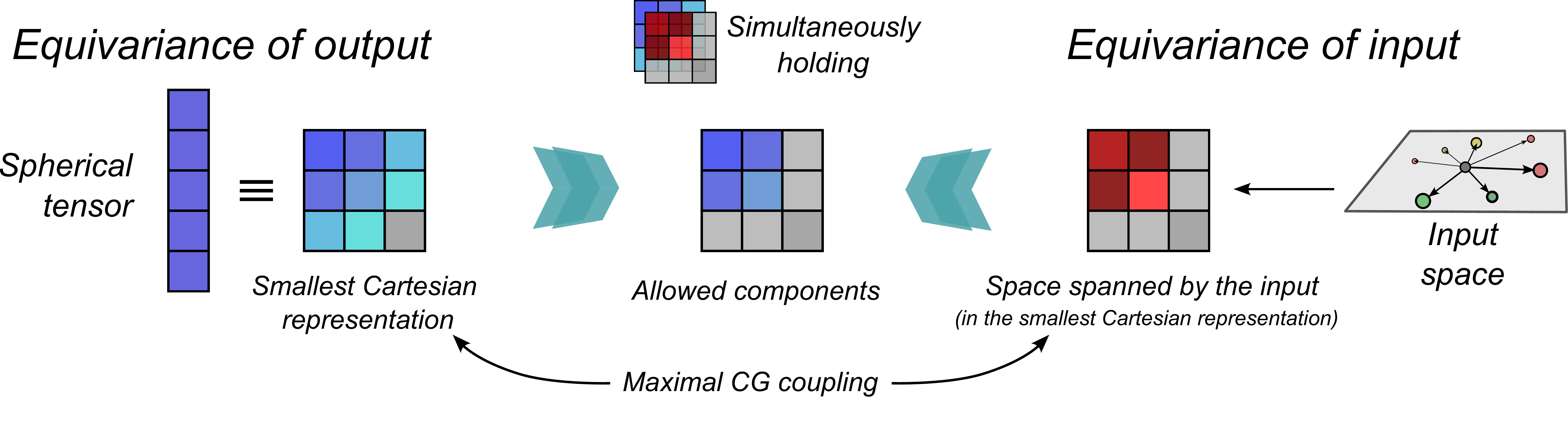}
    \caption{Graphical representation of the result from Prop.~\ref{prop1}. A spherical tensor (on the left) of order $\lambda$, can be always encoded into a Cartesian tensor of rank at least $\lambda$. When the rank and the angular momentum coincide, we have the smallest Cartesian representation that can accommodate for the tensor, realized by the maximal CG coupling. 
    In the figure, a tensor of $\lambda=2$ is encoded in the symmetric and traceless components of a Cartesian tensor of the same rank. 
    Applying the same maximal CG coupling to the vectors in the input space (on the right) provides a representation of the same rank. In the figure, the input generates the xy-plane, and its symmetric and traceless representation of rank $2$ is depicted. 
    The main result (converging to the center) is that, to ensure the equivariance condition of Eq.~\eqref{eq:equivariance}, we can consider only components of the tensor in the same space generated by the input, while the other must vanish.}
    \label{fig:proof_meaning}
\end{figure*}

We can now state our first result:
\begin{prop}\label{prop1}
    If $\bm T_\lambda$ is a proper harmonic tensor of the vector inputs $\{\br_i\}_{i=1}^n$, then it must lie in the subspace spanned by the maximal CG coupling $(\br_{i_1}\ocg\ldots \ocg \br_{i_\lambda})_\lambda$, with $i_j \in \{1,\ldots, n\}$, $\forall j$.
\end{prop}

This proposition is a direct generalization of the results of Ref.~\citenum{NEURIPS2021_f1b07759} to harmonic tensors, and a graphical representation is depicted in \autoref{fig:proof_meaning}. The detailed proof is reported in the SI, and we give here the general outline.

The starting point lies in the observation that a Cartesian tensor of rank $r$ contains spherical components of angular momentum up to the rank: this means that the smallest Cartesian tensor that can contain a spherical tensor of rank $\lambda$ is the one for which the rank is itself $\lambda$.
This procedure is realized by means of the maximal coupling given in Eq.~\eqref{eq:maximal_coupling}. 
In other words, the components of a spherical tensor can be given with respect to a basis obtained by maximal couplings only (for example, all possible maximal couplings of vectors of the canonical basis). 
The second observation is that any general vector $\bm v\in \mathbb{R}^3$ can be written as $\bm v = \bm u +\bm w$, where $\bm u$ is a vector in the span of the input $\bm u \in \mS := \text{span}(\{\br_i\})$, and $\bm w \in \mS_{\perp}$ is a vector in its orthogonal complement.
Thus, any element of the basis of choice can be written as linear combinations of maximally contracted vectors in $\mS$ and in its orthogonal complement $\mS_\perp$ only. As the angular momentum of the tensor is $\lambda$, we are maximally contracting  $\lambda$-length tuples of vectors at a time.
We can separate all the terms containing an even number of vectors in $\mS_\perp$ from the ones containing an odd number. Thus, we get the partition
\begin{equation}\label{eq:explain_proof1}
    \bm T_{\lambda} (\{\br_i\}) = \bm T^{\text{even}}_\lambda(\{\br_i\}) + \bm T^{\text{odd}}_\lambda(\{\br_i\}).
\end{equation}
We can then apply a transformation $Q\in O(3)$ such that $\bm Q^{(1)}\bm u= \bm u$, for all $\bm u \in \mS$ and $\bm Q^{(1)} \bm w = -\bm w $, for all $\bm w \in \mS_\perp$. 
As any vector of the input $\br_i$ is trivially in $\mS$, then we have that by the equivariance conditions of Eq.~\eqref{eq:equivariance} the tensor is unchanged, namely $\bm T_\lambda (\{\bm Q^{(1)}\br_i\})= \bm T_\lambda(\{\br_i\})$. 
Since the maximal contractions are linear, any contraction containing an even number of vectors in $\mS_\perp$ is also left unchanged, as we are changing sign an even number of times:
Thus, also $T_\lambda^\text{even}$ remains unchanged.
On the contrary, $T_\lambda^\text{odd}$ acquires a sign. In practice we have the following chain of equivalencies
\begin{equation}
\begin{split}
    \bm T_\lambda(\{\br_i\}) = \bm T_\lambda(\{\bm Q^{(1)}\br_i\}) = \bm Q^{(1)}\bm T_\lambda(\{\br_i\}) 
    \\
    = \bm T^{\text{even}}_\lambda(\{\br_i\}) - \bm T^{\text{odd}}_\lambda(\{\br_i\}).
\end{split}
\end{equation}
Comparing this result with Eq.~\eqref{eq:explain_proof1}, we deduce that $\bm T_\lambda^{\text{odd}}=0$ and we are left with only an even number of vectors $\bm v\in \mS_\perp$.
We can consider three cases separately, with respect to the possible size of $\mS$: if $\mS = \mathbb{R}^3$, then we have nothing else to prove. If, instead, the inputs generate a plane, $\dim(\mS)=2$, then we have only one vector $\bm p$ generating $\mS_\perp$.
Thus, $\bm T_\lambda^{\text{even}}$ contains terms like $(\bm v\ocg \bm v)_2$: as shown in the example of Eq.~\eqref{eq:example_projection_orthogonal}, these terms lie entirely in the space generated by vectors orthogonal to $\bm v$, again leading to the proof.
In the extreme case in which all the inputs are aligned, when $\dim(\mS)=1$, the proposition is proved by noticing that all the components of an equivariant tensor vanish but for the one with $\mu = 0$ (cylindrically symmetric). 
Thus the proof is concluded by showing that, on the contrary, contributions from $\mS_\perp$ introduce non-vanishing components with $\mu\neq0$.

An immediate consequence of Prop.~\ref{prop1} is the corollary:
\begin{corollary}\label{corollary2}
    If $\bm T_\lambda$ is a proper harmonic tensor of the vector inputs $\{\br_i\}_{i=1}^n$, then it is possible to write
    \begin{equation}
        \bm T_{\lambda}(\{\br_i\}) = \sum_{i_1\le\ldots \le i_\lambda} f_{i_1\ldots i_\lambda}(\{\br_i\}) (\br_{i_1}\ocg \ldots \ocg \br_{i_\lambda})_{\lambda},\label{eq:harmonic-expansion-1stq}
    \end{equation}\\
    with $f_{i_1\ldots i_\lambda}(\{\br_i\})$ scalar functions or, equivalently,
    \begin{equation}
        \bm T_{\lambda}(\{\br_i\}) = \sum_{\substack{l_1,\ldots, l_n = 0\\l_1+\ldots+ l_n = \lambda}}^\lambda \phi_{l_1\ldots l_n}(\{\br_i\}) (\br_{1}^{\ocg l_1}\ocg \ldots \ocg \br_{n}^{\ocg l_n})_{\lambda},\label{eq:harmonic-expansion-2stq}
    \end{equation}
    with $\phi_{l_1\ldots l_n}(\{\br_i\})$ scalar functions.
\end{corollary}

We mention that we can choose an ordering for the vectors in the sum of Eq.~\eqref{eq:harmonic-expansion-1stq} because of the commutativity of the maximal CG coupling.
The corollary shows how the ``geometric'' part of the tensor, responsible for the equivariant behavior, can be fully encoded into maximal coupling of the inputs, modulated by scalar functions. 
In particular, in the second half of the corollary, the sum is constrained to values of the $l$ channels such that their sum is equal to $\lambda$: this is in stark contrast with other general formulations (see, for example, Refs.~\citenum{drau19prb,will+19jcp,niga+20jcp}) in which the contractions reach arbitrary large channels for the coupling of angular momenta. 
On the contrary, the formulation above shows that it is possible to consider only the \emph{smallest} possible coupling of the input vector which is compatible with the target angular channel.
Still, the corollary above does not provide any regularity property for the scalar functions. This is addressed in the following Proposition:

\begin{prop}\label{proposition3}
    If $\bm T_\lambda$ is a proper harmonic tensor, that can be expressed in terms of polynomials of the inputs $\{\br_i\}_{i=1}^n$, then the scalar functions of its expansion, $f_{i_1\ldots i_\lambda}(\{\br_i\})$ and $\phi_{l_1\ldots l_n}(\{\br_i\})$, can be chosen to be polynomial.
\end{prop}
This proposition provides the link to ML architecture: in particular, with the equivariant (geometric) part completely characterized by the maximal coupling of the input vectors, the specific description of the tensor $\bm T_\lambda$ is transferred to the scalar functions, which can now be approximated with some architecture of choice. 
This is closely related to the approach proposed in Ref.~\citenum{will+19jcp}, where $\lambda-$SOAP descriptors were enhanced with scalar representation for the atomic environment.
However, while the coupling in the $\lambda$-SOAP there was left unconstrained, here we consider only the maximal coupling of the vectors. 
We will discuss and converge to the $\lambda$-SOAP representation in the next section, where we will propose a practical approximation for the expressions above.

We conclude this section by providing similar results for the case of pseudotensors. 
In this case, naively employing the maximal coupling cannot lead to the correct result, as it always produces a proper tensor.
This can be addressed by defining a \emph{pseudo-maximal} coupling as a maximal coupling which contains one, and only one, pseudovector, as shown in the SI. 
This observation is now enough to generalize the previous results and obtain the Proposition:

\begin{figure*}[t]
    \centering
    \includegraphics[width=.95\textwidth]{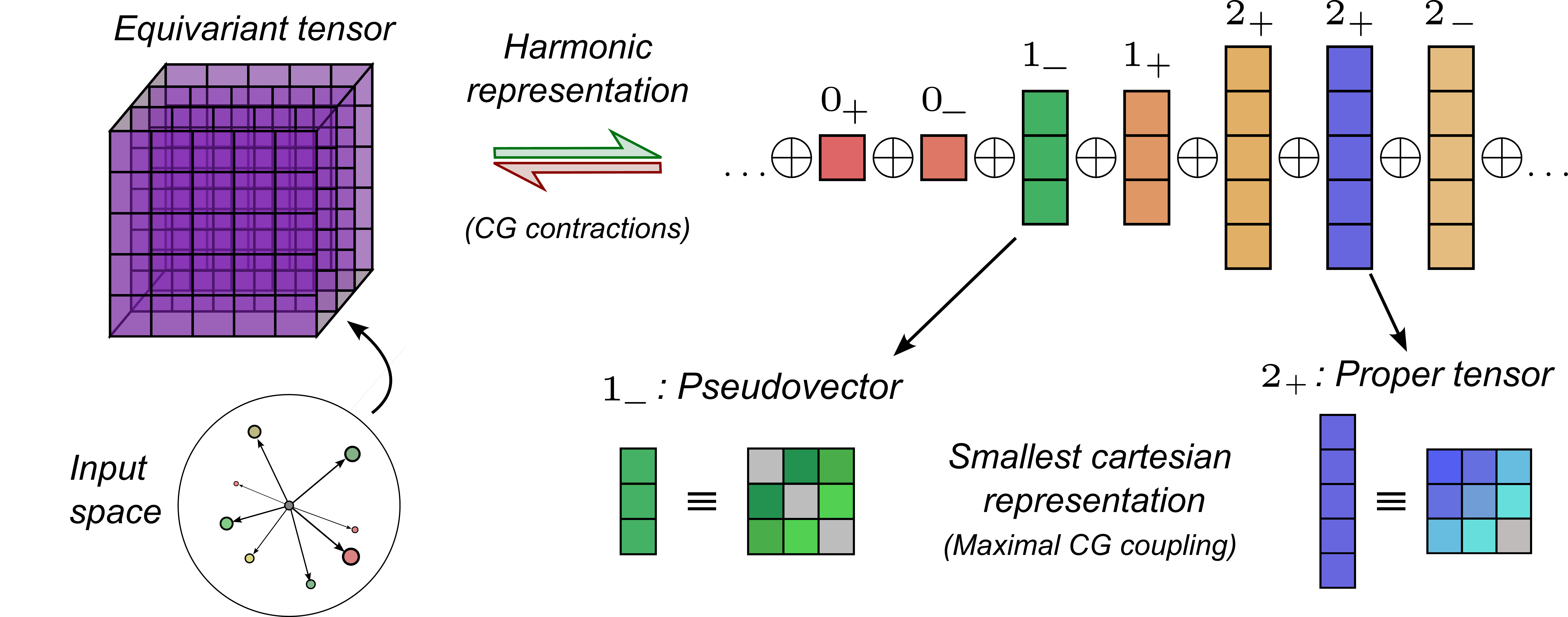}
    \caption{From left to right: let us consider a Cartesian tensor of arbitrary rank which is equivariant with respect to some input vectors $\{\br_i\}$.
    Such tensor admits a decomposition in its harmonic components by means of CG contractions: the result is a collection of irreducible representations with respect to the action of elements of the group $O(3)$. 
    In particular, the notation $l_{\pm}$ indicates the angular momentum (dimensionality) of the representation and its behavior under inversion, proper in the case $l_+$ and pseudo for $l_-$, as per Eq.~\eqref{eq:parity}.
    Then, we can apply the results of \autoref{sec:methods} on each of the irreducible components: in particular, the main idea behind the theoretical results is to encode each of these representations into the smallest Cartesian tensor that can contain them, and this leads directly to the representation in terms of the maximal coupling. 
    For example, in the figure it is shown how a pseudovector $1_-$ can be encoded into the antisymmetric part of rank 2 Cartesian tensor, while a proper $2_+$ tensor can be encoded in the symmetric and traceless part.
    Since the CG contractions from the original tensors and its spherical components is invertible, after applying the results of \autoref{sec:methods}, one can go back to a representation of the full Cartesian tensor in terms of maximal couplings only.}
    \label{fig:main_idea}
\end{figure*}

\begin{prop}\label{prop4}
    If $\bm \Theta_\lambda$ is a harmonic pseudotensor of the vector inputs $\{\br_i\}_{i=1}^n$, then it must lie in the subspace spanned by the pseudo-maximal CG coupling $((\br_{i_0}\ocg\br_{i_1})_1\ocg \br_{i_2}\ldots \ocg \br_{i_{\lambda}})_\lambda$, with $i_j \in \{1,\ldots, n\}$, $\forall j$.
\end{prop}
The complete proof of this and the following results is reported in the SI.
Also the following Corollary is a generalization of the previous one:

\begin{corollary}\label{corollary5}
    If $\bm \Theta_\lambda$ is a pseudotensor function of the vector inputs $\{\br_i\}_{i=1}^n$, then it is possible to write
    \begin{equation}\label{eq:expansion_pseudotensors}
    \begin{split}
        &\bm \Theta_{\lambda}(\{\br_i\}) 
        \\
        &= \mathrm{i}\sum_{i_0>i_1} \sum_{i_2\geq\ldots\geq i_\lambda} f_{i_0\ldots i_{\lambda}}(\{\br_i\}) \big((\br_{i_0}\ocg\br_{i_1})_1\ocg \br_{i_2}\ldots \ocg \br_{i_{\lambda}}\big)_\lambda,
    \end{split}
    \end{equation}\\
    with $f_{i_0\ldots i_\lambda}(\{\br_i\})$ real scalar functions.
\end{corollary}
In this work we adopt a formalism for which the pseudotensorial components are given in terms of real numbers, as this is more memory friendly than complex numbers in terms of practical implementation. 
This is achieved by the imaginary unit in the equation above as we recall that it holds that $(\br_{i_0} \ocg \br_{i_1})_{1} = -\mathrm{i} (\br_{i_0} \times \br_{i_1})/\sqrt{2}$: 
in this way, taking the functions $f_{i_0\ldots i_{\lambda}}$ to be real, the components are real.
We mention that an approximation analogous to the one of Corollary~\ref{corollary2} in terms of $\phi$ functions is trivially achievable also in this case.
However, as it would involve considering also all the possible pairs of input vectors, it is not as useful and thus is not presented here.
Finally, we are able to make the results more concrete for practical application by the following Proposition

\begin{prop}\label{proposition6}
    If $\bm \Theta_\lambda$ is a harmonic pseudotensor that can be expressed in terms of polynomials of the inputs $\{\br_i\}_{i=1}^n$, then the scalar functions of its expansion, $f_{i_0\ldots i_{\lambda}}(\{\br_i\})$ can be chosen to be polynomial.
\end{prop}
As in the case of proper tensors, this result allows to use these results into a practical framework, for example making the scalar function $f_{i_1\ldots i_\lambda}$ be the output of a deep architecture.
In both cases, one can see clearly that these results, as stated, rapidly become impractical as the number of vectors  in the input and/or the tensor order $\lambda$ grow. 
The next section will address this problem and introduce our model.

\subsection{An example}\label{sec:coupling_vectors_results}
Applying our results to a tensor obtained by the coupling of vectors (the most trivial example of polynomial harmonic (pseudo)tensors), allows one to obtain identities like the one of Eq.~\eqref{eq:fundamental_identity} for virtually any coupling and any angular momentum.
Perhaps, the best way to show this is by means of an example. 

Let us take the following coupling of six arbitrary vectors
\begin{equation}
\begin{split}
&\Big(\big(\big(\big((\br_1 \ocg \br_2)_2 \ocg \br_3\big)_2\ocg \br_4\big)_2\ocg \br_5\big)_3\ocg \br_6\Big)_4
\\
&= \sum_{i_1 i_2 i_3 i_4} f_{i_1 i_2 i_3 i_4}(\{\br_i\}) (\br_{i_1}\ocg \br_{i_2} \ocg \br_{i_3} \ocg \br_{i_4})_4.
\end{split}
\end{equation}
As the proposed coupling leads to a proper tensor, we can directly apply Prop.~\ref{prop1}, Coroll.~\ref{corollary2} and Prop.~\ref{proposition3}, which leads to the right-hand side.
By observing that the terms in the left-hand side are homogeneous of degree 1 in each of the vectors, we further observe that all the indexes in the summation must be different (the addends are identically zero if two or more indexes are equal) and that the function $f_{i_1 i_2 i_3 i_4}$ can only be of the form 
$$
f_{i_1i_2i_3i_4} = \alpha_{i_1i_2i_3 i_4} (\br_{i_5}\cdot \br_{i_6}),
$$
for some real numbers $\alpha_{i_1i_2i_3i_4}$, and where all the vector indexes are different. We can compare these results with Eq.~\eqref{eq:fundamental_identity}, and remark that this example can be trivially generalized to any CG contraction of any arbitrary number of vectors. 
In this way we just obtained the new result that Eq.~\eqref{eq:fundamental_identity} is a special case of a huge plethora of identities, that exist for any angular momentum, which all lead to expressions written in terms of the maximal coupling. 
In other words, one can always find an explicit recoupling that allows writing the contractions in terms of (pseudo-)maximal couplings and scalar functions only.

\subsection{Decomposing an arbitrary Cartesian tensor}

Even though we have focused our discussion on the description of irreducible spherical tensors, our construction can be applied rather straightforwardly to any equivariant Cartesian tensor, $\bm T(\{\br_i\})$, which depends on the input positions $\{\br_i\}_{i=1}^n$. 
Our strategy is depicted in \autoref{fig:main_idea}: the first step is to extract the harmonic representation of the tensor. 
This is done by means of CG contractions and gives the irreducible decomposition of the tensor with respect to its behavior under the action of rotations and inversions (we report the details of this decomposition in the SI).
In particular, we are able to reach representations, at most, of the rank of the tensor $\bm T$. 
Each of the terms of the decomposition can then be expanded in terms of the maximal coupling of the inputs (or pseudo-maximal in the case of pseudotensor terms). 
Given that the original CG decomposition is an operation that can be inverted, we are then able to perform the inverse contraction and reconstruct the original tensor, completing the recipe to describe Cartesian tensors.

As we already mention, another work that addresses the expansion of a Cartesian tensor is the one from Ref.~\citenum{gregory2024learningequivarianttensorfunctions} where an expansion in terms of external products and additional Kronecker-delta tensors has been achieved. 
The results there show that the naive external products are not able to cover the whole space of the equivariant tensor, and explicitly point to what the ``missing'' terms are.
However, our results are fundamentally different: 
firstly and foremost, an expansion in terms of irreducible components requires only maximally coupled expansion, without the deficiencies of a basis in terms of pure external product basis. 
Moreover, contrary to the external products, the maximal coupling is totally symmetric, allowing for a substantial reduction of the number of scalar functions that have to be considered (see, for example, Eq.~\eqref{eq:harmonic-expansion-1stq}). 
The two results are also mathematically separated: trying to directly obtain the spherical components from  external products requires the use of all the coupling paths and not only the maximally coupled ones.
On the contrary, unfolding a maximally coupled CG contraction (going back from the spherical representation to the Cartesian tensor as in Fig.~\autoref{fig:main_idea}) does not lead to an explicit expression in terms of external products. 
We report a direct example of this difference in the SI, where an intuitive explanation on the presence of the Kronecker tensors in Ref.~\citenum{gregory2024learningequivarianttensorfunctions} and the lack thereof in our formalism is explained in terms of extraction of scalar subspaces.
We also mention how the use of a spherical representation has few advantages on its own.
Most tensorial quantities of interests possess some symmetry, and a spherical decomposition naturally mirrors the symmetries at play, since the symmetry-breaking irreducible representations are forced to vanish.
Leveraging on this, a spherical representation allows also for a reduction on the number of targets that have to be considered, allowing for slimmer and less redundant models.

\subsection{Permutational invariance}\label{sec:permut_invariance}

As we already mentioned for Eq.~\eqref{eq:permut_inv_scalars_universal}, the number of scalar functions is greatly reduced when the covariant target is permutation invariant with respect to the input. 
Similar simplifications holds also for the expression of Coroll.~\ref{corollary2} and Coroll.~\ref{corollary5}. 
However, the commutativity of the maximal coupling plays a crucial role here and so the analogous generalization for our results is better introduced by a direct example: for the case of a $\lambda = 3$ tensor, which is permutation invariant with respect to its inputs, it reads
\begin{equation}\label{eq:example_perm_invariance3}
\begin{split}
    \bm T_{3}& (\{\br_i\}) = \sum_{i_1} f_0 (\br_{i_1}, [\br_j]_{j\neq {i_1}}))\br_{i_1}^{\ocg 3} 
    \\
    &+ \sum_{i_1 i_2} f_1 \big(\br_{i_1}, \br_{i_2}, [\br_j]_{j\notin \{ i_1, i_2\}}\big)\big(\br_{i_1}^{\ocg 2}\ocg \br_{i_2}\big)_3
    \\
    &+ \sum_{i_1 i_2 i_3} f_2 \big([\br_{i_1}, \br_{i_2}, \br_{i_3}], [\br_j]_{j\notin \{ i_1, i_2, i_3\}}\big)\big(\br_{i_1}\ocg \br_{i_2}\ocg \br_{i_3}\big)_3,
\end{split}
\end{equation}
where the square brackets indicate that the functions are permutationally invariant with respect to any swap of the enclosed vectors. 
Here we can appreciate how the number of scalar functions is dramatically reduced to only three. 
This is the number of ways in which we can reach the angular momentum of $\lambda = 3$ by means of the maximal coupling of vectors, and accounting for the fact that the same vector can appear multiple times in one maximal coupling. 
The generalization of this approach is straightforward, albeit tedious, and is reported in the SI.
Moreover, because in atomistic models it is crucial to enforce permutational invariance with respect to atoms belonging to the same species (or, more abstractly, colors~\cite{Hartmut2024}) then the representation becomes $\bm T_{\lambda}\equiv \bm T_{\lambda}(\{z_i,\br_i\})$, where each of the position arguments of the scalar functions in Eq.~\eqref{eq:example_perm_invariance3} is supplemented by the corresponding atomic ``color'' $z_i$.

\section{Make it practical: the $\lambda$-MCoV model}\label{sec:practical}

\subsection{Atom-centered reference frames}\label{sec:3A}

As already mentioned, the results from Prop.~\ref{corollary2} and Prop.~\ref{corollary5} are impractical due to the scaling with respect to the number of atoms, even in the simplification offered by permutation invariant symmetry. 
Two separate problems arise: how to simplify the expression to have more favorable scaling and how to describe the scalar functions. 
We will now address the former, postponing the discussion on the latter to Sec.~\ref{sec:Tensor_basis}.

Starting from Eq.~\eqref{eq:harmonic-expansion-1stq} we can see how the number of scalar function is unsustainable for any practical purposes. Indeed, for an harmonic tensor of degree $\lambda$ and for $n$ vectors in the input space, this number is $n^\lambda$ for proper tensors. An almost identical observation can be done from Eq.~\eqref{eq:expansion_pseudotensors} for the case of pseudotensors.
As pointed out in \autoref{sec:permut_invariance}, permutational invariance can provide a significant reduction of these numbers, making it more manageable even for relatively large $n$ and $\lambda$. 
However, while this reduction is most effective when all the vectors belong to the same color, the scaling problem is always present for terms that include two or more vectors belonging to different colors.
Even more, to fully use permutational invariance, one has to sacrifice generality and provide ad hoc expressions for all possible combinations of different colors. 
In the following, we will introduce and employ a different strategy, consisting of using only three effective vectors: starting again from Eq.~\eqref{eq:harmonic-expansion-1stq} we can momentarily assume that we can cherry-pick three vectors in the input $\{\bm q_\alpha\}_{\alpha=1}^3 \subset \{\br_i\}_{i=1}^n$, such that they span the full space $\mS$ (which we recall being the space generated by the input vectors, $\mS= \text{span}(\{\br_i\})$). 
Leveraging the linearity of the maximal coupling, we can further assume to write all the vectors $\{\br_i\}$ of the input as a linear combination of these three vectors only.
Clearly, this procedure is not well-defined globally: for example, the vectors $\{\bm q_\alpha\}$ can continuously become coplanar (or even aligned) while the space $\mS$ remains unchanged.
However, for vectors for which these assumptions hold, using the results of Eq.~\eqref{eq:fundamental_identity} and Corollary~\ref{corollary2}, we obtain the following general expressions for a tensor $\bm T_\lambda$, separated by values of the angular momentum $\lambda$. 
For $\lambda=1$, we get
\begin{equation}\label{eq:Q_lambda1}
\bm T_{1}(\{\br_i\}) = \sum_{\alpha = 1}^3 f_{\alpha} (\{\br_i\})\,\bq_{\alpha}
\end{equation}
The expression for $\lambda =2 $ is obtained by summing the two terms:
\begin{equation}\label{eq:Q_lambda2}
\begin{split}
\bm T_{2}(\{\br_i\}) &= \sum_{\alpha = 1}^2 f_\alpha(\{\br_i\}) \bq_\alpha^{\ocg 2}
\\
&\quad+ \sum_{\substack{\alpha_1,\alpha_2= 1\\\alpha_1<\alpha_2}}^{3} g_{\alpha_1\alpha_2}(\{\br_i\})\big(\bq_{\alpha_1}\ocg \bq_{\alpha_2}\big)_{2},
\end{split}
\end{equation}
where we notice that, by applying Eq.~\eqref{eq:fundamental_identity}, the first sum does not contain $\bm q_3$. The more general expression for $\lambda\geq 3$ is
\begin{equation}\label{eq:Q_lambda_gen}
\begin{split}
    \bm T_{\lambda} (\{\br_i\}) &= \sum_{l= 0}^\lambda f_{l} (\{\br_i\}) \Big(\bq_1^{\ocg l}\ocg {\bq}_2^{\ocg(\lambda-l)}\Big)_{\lambda}
    \\
    &\quad+\sum_{l=0}^{\lambda-1} g_{l} (\{\br_i\})\Big(\bq^{\ocg l}_1\ocg \bq_2^{\ocg (\lambda-l-1)}\ocg \bq_3\Big)_{\lambda\mu}.
\end{split}
\end{equation}
We recall that the self-maximal coupling, $\bq_\alpha^{\ocg l}$, is proportional to the solid spherical harmonics, namely $\bq_\alpha^{\ocg l} \propto \abs{\bq_\alpha}^l \bm Y_{l} (\hat{\bq}_\alpha)$ (here $\bm Y_l = (Y_{lm}: \abs{m} \leq l)$ is the vector of spherical harmonics);
practically, we will use only solid spherical harmonics in our architecture, letting the scalar functions absorb the unessential proportionality constants.
The derivation of Eqs.~\eqref{eq:Q_lambda1}, \eqref{eq:Q_lambda2} and \eqref{eq:Q_lambda_gen} is reported in the SI, but we mention that they have been obtained by assuming that the vectors are ordered such that, where possible, $\bq_1$ and $\bq_2$ are linearly independent.
An important observation is that the number of scalar functions is always $2\lambda + 1$, in accord with the degrees of freedom of the equivariant tensors. 
While this is not surprising, given that three vectors can also constitute a basis for 3D space, this ensures that we reached the smallest possible representation for the general tensor $\bm T_\lambda$.
In other words, we shifted the descriptive burden into the scalar functions (which can possess any non-linearity) while ensuring that we have enough scalar terms to describe all the degrees of freedom of an equivariant object. 
A minor detail in this representation is that we are allowing the scalar functions to directly take the vectors $\{\br_i\}$ as input, namely they are not expanded in terms of the $\{\bq_\alpha\}_{\alpha=1}^3$.
While the above procedure directly leads to the minimal number of scalar function that can be, in general, utilized (corresponding to the number of free components of a tensor), we already noticed how singling out a frame of reference is not globally possible: 
not only can we not always address the full space of $\mS$, but it would also inconsistent with permutational invariance of identical position. 
We mention that the analogous results for pseudotensors are reported in the SI, which also show a separation in terms that depend on pseudoscalars and cross products.

\subsection{Frame averaging}

A possible strategy to obtain a practical universal approximator based on the results of the three-vector expansions is given by Refs.~\citenum{pozd-ceri23nips} and~\citenum{niga+24aplml}. 
The main idea is to take an average over expressions obtained by taking into account all possible coordinate systems built from the point cloud.
This approach can be applied also to a mean of expressions-like Eqs.~\eqref{eq:Q_lambda1}, ~\eqref{eq:Q_lambda2}, and~\eqref{eq:Q_lambda_gen}, constructed over all possible triplets of vectors. On the one hand, as these expressions are easily separable in terms that depend on one, two or three positions at the time, their average can be separated and made more practical. 
On the other hand, each of the terms of the mean will contribute with its own scalar functions: we will have to consider $2\lambda+1$ scalar functions for each of the possible triplets of atomic species (in the presence of permutational invariance).   
Moreover, even if Refs.~\citenum{pozd-ceri23nips} and~\citenum{niga+24aplml} use the clever consideration that a frame of reference can be defined by just a pair of vectors, reducing the scaling with respect to the number of positions from cubic to quadratic, this is done by considering cross products. 
Unfortunately, this is not directly applicable in this context, as all vectors in the expansions above must be proper (the cross product is pseudo), and we must ensure that all three vectors lie in the span $\mS$ (a cross product can go outside of the span). 
Therefore, while a frame-averaging strategy is surely applicable, especially in the context of small point clouds, the bad scaling with respect to the number of positions is still rather poor, and therefore we will pursue a more efficient, albeit approximated, approach.

\subsection{The Maximally Coupled Vector (MCoV) model}

A usual route to target tensorial quantities is to use an expansion written in terms CG couplings of spherical expansions~\cite{bart+13prb}. 
As these constitute a permutational invariant basis~\cite{drau19prb, niga+20jcp, duss+22jcp, Hartmut2024} one can increase the accuracy with an increase of body-order and angular channels involved. 
While using spherical couplings is complete, one has to perform all possible CG contractions compatible with the target tensors, which is contrary to the simplification provided by the maximal couplings.

Our approximation consists of using directly Eq.~\eqref{eq:Q_lambda1}, \eqref{eq:Q_lambda2} and \eqref{eq:Q_lambda_gen} with the following expression for the vectors $\bq_\alpha$:
\begin{equation}\label{eq:q_vectors_from_spherical_exp1}
    \bq_{\alpha}([\br_{i}]) = \sum_{zn} W_{\alpha,zn} \bm \rho_{z, n1}([\br_i]) .
\end{equation}
where $W_{\alpha,nz}$ are learnable weights (a linear layer) and where we used the spherical expansion~\cite{duss+22jcp,Hartmut2024,domina2025generalformalismmachinelearningmodels} defined as
\begin{equation}\label{eq:spherical_expansion}
    \bm \rho_{z, nl}([\br_i]_{i\in z}) = \sum_{i \in z} R_{nl} (r_i) \bm Y_{l}(\hat{\br}_i),
\end{equation}
with $l=1$ (the vectorial case). 
Here $\{R_{nl}\}$ is a complete set of radial functions and we partitioned all the positions in terms of identity classes labeled by the atomic species (more generally, color) $z$. 
The expansion is usually assumed to be local. The radial functions are defined inside a sphere of radius $r_{\text{cut}}$ and smoothly vanish to zero approaching this cut-off distance. The linear layer in Eq.~\eqref{eq:q_vectors_from_spherical_exp1} acts on the radial channels and on the species (colors): in the language of ML applications, this step also incorporates a chemical embedding~\cite{will+18pccp} that creates a more favorable scaling with the number of atomic species of the input.

As already mentioned, this strategy is not always applicable, in particular when in presence of highly symmetric configurations of identical positions. We will discuss the limitations and ways to overcome them in Sec.~\ref{sec:limitations-and-correction}.

By plugging Eqs.~\eqref{eq:spherical_expansion} and~\eqref{eq:q_vectors_from_spherical_exp1} into ~\eqref{eq:Q_lambda1}~-~\eqref{eq:Q_lambda_gen}, we obtain a
minimal representation that contains the correct number of degrees of freedom (number of components of the tensor) encoded in as many scalar functions. 
Here, equivariance is obtained by means of maximal couplings only. 
In particular, having a minimal number of scalar functions makes it feasible to describe them by means of non-linear and deep architectures. 
Moreover, the spherical expansion of Eq.~\eqref{eq:spherical_expansion} is permutationally invariant with respect to swaps of positions belonging to the same species.
Then, to enforce permutational invariance of the tensor, it is sufficient to have permutationally invariant scalar functions. 
We call this model the Maximally Coupled Vector (MCoV) model.

\begin{figure*}[t!]
    \centering
    \includegraphics[width=1\textwidth]{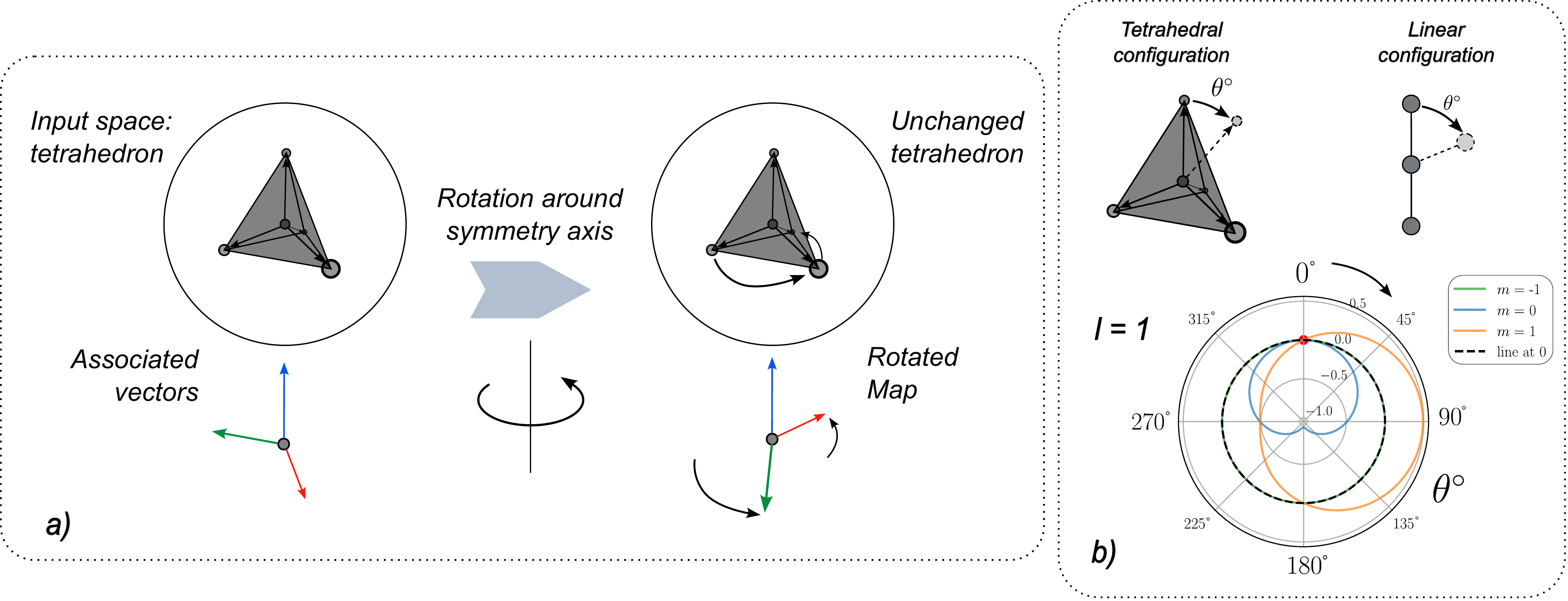}
    \caption{
    a). An example of case for which it is not possible to uniquely associate a frame of reference to the point cloud due to the high symmetry and the permutational invariance. 
    The point cloud is described from the center of mass of the tetrahedron, and any rotation around one of the symmetry axis will leave the points cloud unchanged while rotating any frame associated with it: to guarantee the uniqueness of the map, all the components of the frame in the perpendicular direction to the axis of rotation must vanish.
    Applying the same rationale to any of the symmetry axis we have that the only map compatible with the symmetry operations is the trivial one, with only null vectors.
    \\
    b). Polar plot representing the spherical expansion of Eq.~\eqref{eq:spherical_expansion} for $l=1$ as the point at the top of the tetrahedron traces a circle in the $xz$-plane. 
    The $0^\circ$ position represent the fully symmetric configuration: all the components of the spherical expansion go to zero continuously (the red dot, intercepting the dashed black line) in the fully tetrahedral configuration, in accordance to a). 
    The very same spherical expansion for $l=1$ is found also in the linear configuration, with the spherical expansion vanishing in the aligned positions. 
    The linear case will be investigated in a real scenario for molecules of CO\textsubscript{2} in \autoref{sec:results_CO2}.
    }
    \label{fig:vectors_not_enough}
\end{figure*}

\subsection{Limitations of the vectors approach and its correction}\label{sec:limitations-and-correction}

This section is devoted to address the limitation of the simplified description of Eqs.~\eqref{eq:Q_lambda1}, \eqref{eq:Q_lambda2} and \eqref{eq:Q_lambda_gen} in the scheme offered by Eq.~\eqref{eq:q_vectors_from_spherical_exp1}.
There are two main limits of this scheme, the first being in the case of highly symmetric configuration (see also Ref.~\citenum{dym2024}) and the second being when the vectors are not able to cover the whole space on which the tensor lives. We will now address both cases, starting from the former. 

Our scheme is equivalent to associating a frame of reference to the point cloud. This is not always possible, in particular when combined with permutational invariance. 
An obvious case in which this approach fails is when Eq.~\eqref{eq:spherical_expansion} produces only null vectors, for example in the case in which the positions in the point cloud form a highly symmetric configuration.
This is shown in \autoref{fig:vectors_not_enough}: it is not possible to define a (non trivial) map between the atomic positions and a harmonic tensor if a vectorial representation does not satisfy all the symmetry operation of the point cloud.

Importantly, we remark that these issues appear only in presence of permutational invariance: if it is possible to label the different positions in the point cloud then it is also possible to define a frame of reference, which also allows us to exactly adopt a 3-vectors framework.

To solve the issue with highly symmetric configurations we follow the same idea of Ref.~\citenum{NguyenLunghi2022}, where the equivariance of tensors is described by directly using the spherical expansion of order $\lambda$, $\bm \rho_{z,\lambda}$ (with no radial channel). 
Thus, we add an augmentation term
\begin{equation}\label{eq:Q_corr}
    \bm T_\lambda^{(\text{corr})}(\{\br_i\}) = \sum_{\beta = 1}^{2\lambda+1 } h_\beta(\{\br_i\}) \sum_{zn} W^{(\text{corr})}_{\beta,zn} \bm \rho_{z, n\lambda} (\{\br_i\}).
\end{equation}
Here, we introduce $2\lambda+1$ scalar functions $h_\beta(\{\br_i\})$, to be able to accommodate for all the possible degrees of freedom of the tensor. Similarly to Eq.~\eqref{eq:q_vectors_from_spherical_exp1} we mix the radial channels and the atomic species by means of the linear layer with weights $W^{(\text{corr})}_{nz}$. 
As when $\lambda=1$ there is no difference between this and Eq.~\eqref{eq:Q_lambda1}, we define $\bm T^{(\text{corr})}_1(\{\br_i\}) = 0$ to avoid redundancies.
An important difference with Ref.~\citenum{NguyenLunghi2022} consists in the use of the radial functions to increment the descriptivity of the representations. 
In this context it is useful to think about the spherical expansion as a $(2\lambda+1)$-dimensional vector: because the harmonic tensor $\bm T_\lambda$ lives in a $(2\lambda+1)$-dimensional space, a general basis should be able to describe all the $(2\lambda+1)$ different directions required to cover the full space. 
From this point of view, different radial basis channels $n$ in the spherical expansion $\bm \rho_{z, n\lambda}$, provide an additional degree of freedom that makes the coverage of more directions possible.
This also shows why the spherical expansion at $\lambda$, alone, would fail more easily than the 3-vector basis of the MCoV model: 
while the $\lambda$-spherical expansion has to cover the full $(2\lambda+1)$ space, the only requirement for the 3 vectors $\{\bq_\alpha\}_{\alpha=1}^3$ is to cover the same space $\mS$ spanned by the input vectors which is at most $3$-dimensional. 
An example is given by the case in which we have only three different distances and one species in our input vectors: on the one hand, the spherical expansion $\bm \rho_{n\lambda}$ can cover only three directions out of the $2\lambda+1$ ones, which causes a severe lack of descriptivity already for fairly small $\lambda$. On the other hand, the $\{\bm q_\alpha\}_{\alpha=1}^3$ vectors are already able to cover the full input space.

We remark that Eq.~\eqref{eq:Q_corr} addresses an almost always local problem: if we have a highly symmetric configuration of the neighbor atoms with respect to a central one (the center of the description), the degeneracy can be resolved when considering another atom that does not exhibit the same symmetry. 
As the global description is usually obtained by a sum over local ones (see Eq.~\eqref{eq:local_atomic_contributions}), then the model usually retains enough descriptive capabilities to overcome the lack of local representations.
This is analogous to what observed in Ref.~\citenum{pozd+20prl} for a different type of representation failure.

The second case of local failure is connected to the lack of descriptivity of all the degrees of freedom. For example, if the atomic positions have all the same distances from the center, then the spherical expansion is independent on the radial channels and points in one direction only, and only one vector can be obtained from it. This lack of descriptivity can be easily addressed by including contractions of spherical expansions in a $\lambda$-SOAP fashion~\cite{gris+18prl}, namely by including also $(\bm \rho_{Z_1, n(l+1)}\ocg \bm \rho_{Z_2, nl})_1$, for positive $l$ values. 
Once again, this would mostly address lifting of local degeneracies, which are not relevant when targeting global quantities. 
From this it can be seen that adding more and more CG contractions makes the model shift from a minimal representation to the complete one provided, for example, by Ref.~\citenum{niga+20jcp}.
Since including further CG contraction in the representation defies the goal of having only a minimal number of couplings at play we will default to an architecture in terms of the vectors of Eq.~\eqref{eq:q_vectors_from_spherical_exp1} and the correction Eq.~\eqref{eq:Q_corr} only, under the assumption that the two bases will rarely fail simultaneously, as they are built from very different constructions. 

\subsection{The architecture: extending the SOAP-BPNN to equivariant targets}\label{sec:Tensor_basis}

\begin{figure*}[t]
    \centering
    \includegraphics[width = \textwidth]{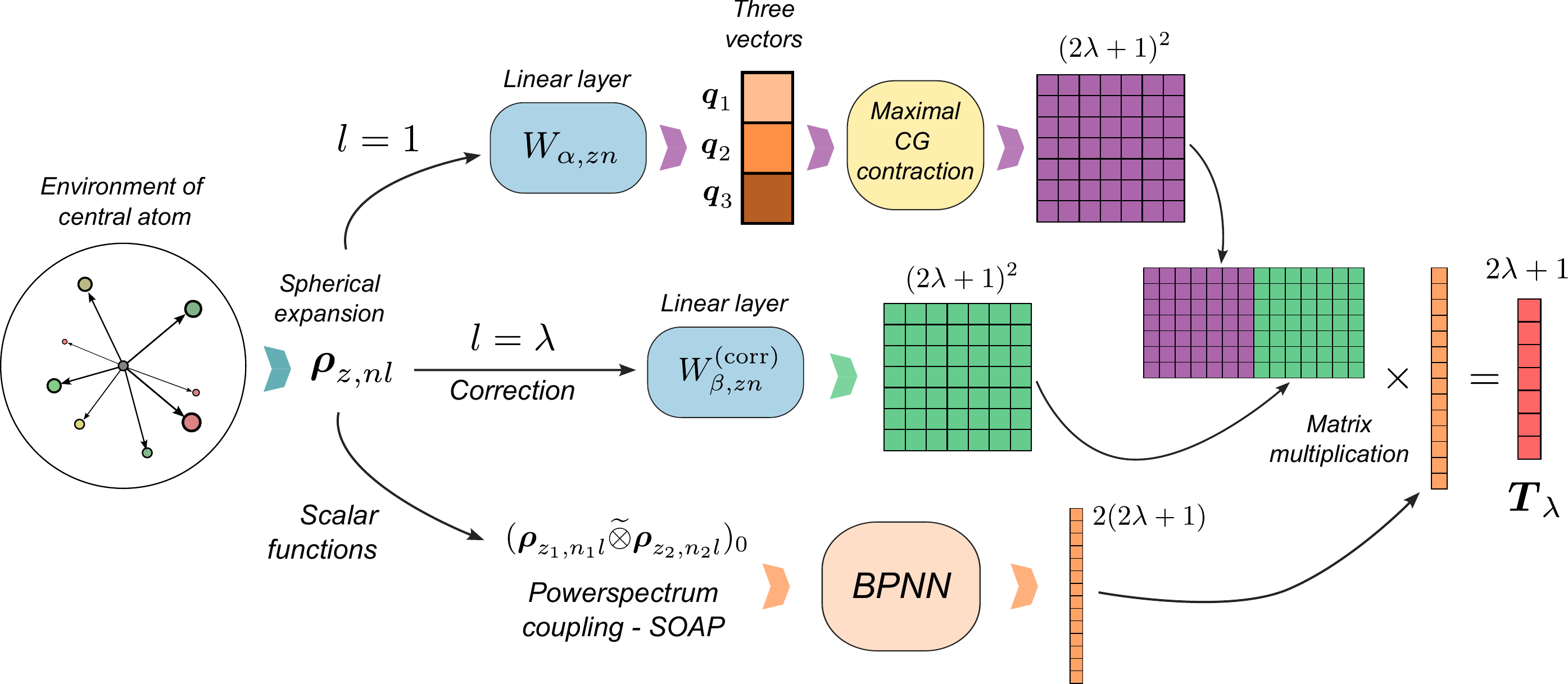}
    \caption{The architecture of the $\lambda$-MCoV. From left to right:
    the input is the local environment centered on one atom (see right-hand side of Eq.~\eqref{eq:local_atomic_contributions}). The relative atomic positions are then used to evaluate the spherical expansion (Eq.~\eqref{eq:spherical_expansion}). The spherical expansions for $l=1$ are then fed into a linear layer that produces the three vectors $\{\bm q_\alpha\}_{\alpha=1}^3$ which are then contracted with maximal coupling according to Eqs.~\eqref{eq:Q_lambda1}-\eqref{eq:Q_lambda2}. This produces a matrix $(2\lambda+1)$ tensors, each in $\mathbb{R}^{2\lambda+1}$. 
    The spherical expansion for $l=\lambda$ are fed into a linear layer producing the same number of tensors, according to Eq.~\eqref{eq:Q_corr}. 
    Finally, the scalar functions are produced by powerspectrum-SOAP contraction, according to Eq.~\eqref{eq:powerspectrum}, and then fed into a Behler-Parrinello Neural Network (BPNN) architecture~\cite{behl-parr07prl} which produces $2(2\lambda+1)$ scalars. 
    These are then contracted with the matrices resulting from the other two paths, resulting in the $2\lambda+1$ harmonic components of the tensor. 
    The weights of the BPNN and $W_{\alpha, zn}$, $W^{(\text{corr})}_{\beta, zn}$ are shared among central atoms of the same species.
    }
    \label{fig:architecture}
\end{figure*}

The prediction of equivariant functions using scalars is particularly convenient, since the computation of expressive scalar representations through an invariant architecture is much simpler and more efficient~\cite{litm+24jcp} than the calculation of expressive equivariant representations, which requires expensive equivariant tensor products~\cite{gris+18prl, niga+20jcp, bata+22nips,niga+22jcp2, batatia2025}. 
It also allows the implementation of non-linear correlations among the input features, in a similar fashion to the architectures outlined in Refs.~\citenum{will+19jcp, gris+19book}.

Historically, in the atomistic domain, the first invariant architecture of this type was the BPNN architecture proposed by~\citet{behl-parr07prl}, still very widely used today despite its known shortcomings Ref.~\citenum{pozd+20prl}. In this work, we will use this architecture to produce the scalar coefficients for the tensor basis of Eqs.~\eqref{eq:Q_lambda1}-\eqref{eq:Q_lambda_gen} and~\eqref{eq:Q_corr}, which will allow us to predict equivariant properties of atomic-scale structures.

In this context, each structure (which is defined by a three-dimensional configuration of atoms) is associated with an equivariant regression target (the energy, the dipole moment, etc.). It is common \cite{gris+18prl,NguyenLunghi2022} to use the assumptions of additivity and locality to predict these quantities from a sum of atomic contributions that only depend on the neighborhood of each atom, up to a cutoff radius $r_\textrm{cut}$. In practice, for global tensors, we assume the partitioning
\begin{equation}\label{eq:local_atomic_contributions}
    \bm T_{\lambda}(\{\br_i\}) = \sum^\text{atoms}_{i} \bm T_{\lambda, z_i}(\{\bm r_{ji}\}),
\end{equation}
where the local tensors $ \bm T_{\lambda, z_i}$ take only inputs with norm smaller than $r_{\text{cut}}$. 
By using only relative vectors, this scheme naturally enforces translational invariance.
Also, the tensors  $ \bm T_{\lambda, z_i}$ depend on the atomic species $z_i$ only, in order to ensure global permutational invariance under swap of identical atoms. 
In practice, this means that the learnable parameters are shared among central atoms belonging to the same species. 
Should the target be an intensive quantity, it does suffice to normalize the expression above to the total number of atoms.

Although the original BPNN uses atom-centered symmetry functions as the scalar descriptors for the neural network, we will use the mathematically equivalent~\cite{bart+17sa} invariant Smooth Overlap of Atomic Positions (SOAP)~\cite{bart+13prb} descriptors as input features. The SOAP descriptors are evaluated by means of the CG contraction on the spherical expansion of Eq.~\eqref{eq:spherical_expansion}, projected on the proper scalar space ($\lambda = 0$). In practice, we will only use the powerspectrum~\cite{bart+13prb} defined as
\begin{equation}\label{eq:powerspectrum}
    p_{z_1z_2, n_1n_2l} := (\bm \rho_{z_1, n_1l}\ocg \bm \rho_{z_2, n_2l})_{0}.
\end{equation}
There is no fundamental reason for the choice of the powerspectrum in place of higher-order correlation terms: if more descriptive features are required, one could overcome the limits of the lack of completeness of the powerspectrum~\cite{pozd+20prl} by taking higher-order correlations of spherical expansions~\cite{drau19prb}.
With these features, the input space is represented by means of the spherical expansions only, which are the only quantities that have to be evaluated from the atomic positions.
The SOAP descriptors are then fed to the BPNN, which is essentially a multi-layer perceptron (one for each atomic species) to compute the final scalar representation. The overall architecture, is shown in \autoref{fig:architecture}, which also illustrates the combination of scalar features and the equivariant basis to predict equivariant spherical tensors.

We call the total model, with the $\lambda$-correction of Eq.~\eqref{eq:Q_corr} and the SOAP-BPNN for the scalar part, the $\lambda$-MCoV model. 

\subsection{Highlighting differences with other architectures}

In this section we will discuss the main differences with previous architecture based on equivariant tensor products. The key distinction is the role of CG contractions. Once the three vectors $\{\bq_\alpha\}$ are determined (e.g., after learning the coefficients of Eq.~\eqref{eq:q_vectors_from_spherical_exp1}), CG contractions are fixed (non-learnable) and used only to form the maximal couplings in~\autoref{sec:3A}. Consequently, expressivity is delegated to the scalar functions, while the maximal couplings do not determine the model depth or its intrinsic body order~\cite{bata+22nips,pozd+20prl,duss+22jcp}.

Another important difference is that CG contractions are typically the bottleneck of hierarchical equivariant models due to (i) the large number of coupling paths compatible with a target, even under modest angular-momentum cutoffs, and (ii) the arbitrariness of the truncation. Our results provide a precise prescription: among all coupling paths, only the maximal one is required to build equivariant objects. This removes arbitrary truncations and most of the computational overhead, retaining only the essential contractions along the maximal path. If desired, hierarchical coupling (e.g., MACE~\cite{bata+22nips} or NICE~\cite{niga+20jcp}) can be applied to the \emph{scalar} part, shifting any remaining overhead from equivariants to scalars; in that case, any CG computational cost would shift from enforcing equivariance to learning invariants.

Finally, the three-vector construction introduces learnable weights on radial channels and atomic species, as in MACE~\cite{bata+22nips} or ALLEGRO~\cite{musa+23ncomm}. These are the only learnable parameters in the equivariant core and preserve equivariance by avoiding mixing across equivariant channels. However, the construction of the $\{\bq_\alpha\}$ vectors requires only the $\ell=1$ spherical expansion, incurring minimal overhead (see Eqs.~\eqref{eq:q_vectors_from_spherical_exp1} and~\eqref{eq:spherical_expansion}).

\begin{figure*}[t]
    \centering
    \includegraphics[width=.9\linewidth]{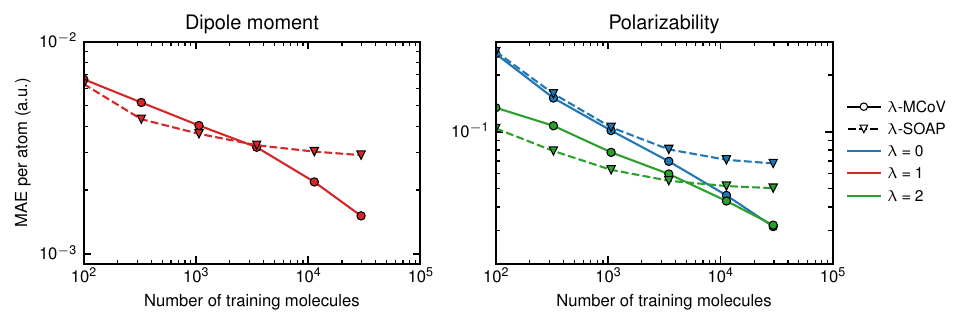}
    \caption{Learning curves obtained from models trained with the proposed architecture compared with linear models on $\lambda$-SOAP descriptors. The training sets are subsets of the QM7-X dataset.}
    \label{fig:tensorbasis vs lambdasoap}
\end{figure*}
\begin{figure*}
    \centering
    \includegraphics[width=.9\linewidth]{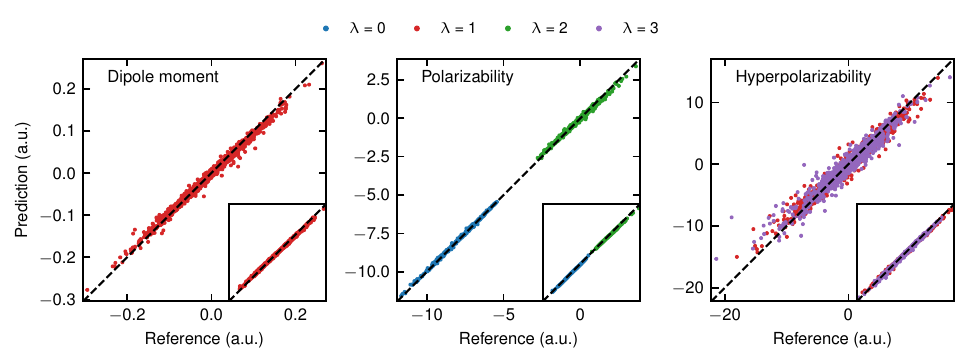}
    \caption{Parity plots of dipole moment, polarizability, and hyperpolarizability per atom of the subset of QM7 used in this work. The main panels show test set properties, while the insets show the training set ones.}
    \label{fig:qm7 multi target}
\end{figure*}

\section{Results}\label{sec:results}

\subsection{Comparison with a simple linear equivariant model}

Computing an expressive representation of scalars, for example from an invariant neural network architecture, is manifestly computationally cheaper than evaluating an equivariant neural network. From a theoretical perspective, the main computational bottleneck of equivariant architectures lies in the repeated CG contractions required to couple irreducible representations. Our proposed architecture largely avoids these costly operations, performing only maximal contractions for equivariant targets and scalar contractions for scalar ones, skipping ttesF.he majority of CG tensor products by substantially truncating the coupling trees. However a case could be made that very simple equivariant models can match the computational efficiency of a scalar-only featurization. In order to explore this comparison, as well as to validate our method, we compare it to a linear $\lambda$-SOAP~\cite{gris+18prl} model in the prediction of dipole moments for the small molecules of the QM7-X dataset~\cite{hoja+21sd}.

\autoref{fig:tensorbasis vs lambdasoap} shows that the more flexible $\lambda$-MCoV, which can reach higher body-orders~\cite{musi+21cr}, not only has comparable accuracy with the $\lambda$-SOAP in the data-poor regime, but it outperforms it in the data-rich regime which, for this example, is around 5000 molecules.

\subsection{Training on multiple equivariant properties: dipole, polarizability, hyperpolarizability}

A distinct advantage of using exclusively scalar functions to predict equivariant properties is that a shared internal representation can be used to predict different equivariant targets, increasing weight-sharing and therefore the ability of the model to re-use geometrical information across separate targets.

To illustrate this possibility, we fit a $\lambda$-MCoV model simultaneously to dipole moments ($\mu$), polarizabilities ($\alpha$) and hyperpolarizabilities ($\beta$) of the $6754$ molecules of the selected subset of QM7~\cite{blum-reym09jacs, rupp+12prl} spanning the CHNO composition space. \autoref{fig:qm7 multi target} shows that all targets are learned to a very good accuracy. 
The MAEs for a 4053\textbar{}1350\textbar{}1351 training\textbar{}validation\textbar{}test split are reported in \autoref{tab:mae-results}.
The target quantities are computed at the mean-field level from Hartree-Fock calculations, exploiting automatic differentiation with \textsc{PySCFAD}~\cite{Zhang2022}, using the \texttt{aug-cc-pvdz} basis set.

\begin{table}[b]
\centering
\small
\begin{tabular}{l
                ccc   ccc   ccc}
\toprule
Split 
  & \multicolumn{3}{c}{$\mu$}
  & \multicolumn{3}{c}{$\alpha$}
  & \multicolumn{3}{c}{$\beta$} \\
\cmidrule(lr){2-4} \cmidrule(lr){5-7} \cmidrule(lr){8-10}
    & $\lambda$ & MAE    & \% 
    & $\lambda$ & MAE    & \%
    & $\lambda$ & MAE    & \% \\
\midrule
Training & 1 & 0.0022 & 5  & 0 & 0.0138 & 1  & 1 & 0.1597 & 5  \\
      &   &        &    & 2 & 0.0215 & 4  & 3 & 0.0942 & 6  \\[0.3em]
Validation   & 1 & 0.0033 & 7  & 0 & 0.0281 & 6  & 1 & 0.3575 & 24 \\
      &   &        &    & 2 & 0.0300 & 6  & 3 & 0.2177 & 14 \\[0.3em]
Test  & 1 & 0.0032 & 7  & 0 & 0.0247 & 5  & 1 & 0.3263 & 22 \\
      &   &        &    & 2 & 0.0283 & 6  & 3 & 0.2144 & 14 \\
\bottomrule
\end{tabular}
\caption{MAE (a.u./atom) and ratio between MAE and training set standard deviation (\%) on the predicted equivariant targets for the QM7 subset.}
\label{tab:mae-results}
\end{table}

\subsection{Application to spectroscopy}

\begin{figure}[h!]
    \centering
    \includegraphics[width=\linewidth]{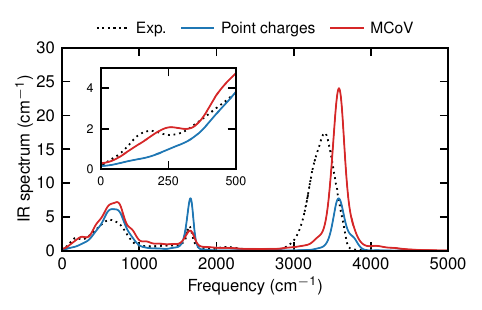}
    \caption{IR spectrum of liquid water at ambient conditions computed from the dipole current autocorrelation function on top of a classical MD trajectory using the q-TIP4P-f forcefield.
    In the inset, a magnification of the peak around 200 cm\textsuperscript{-1}.}
    \label{fig:water spectrum}
\end{figure}

The efficiency of the proposed architecture makes it suitable for applications that require many evaluations of the model on different structures, such as when computing spectra from molecular dynamics (MD) trajectories. 
As a paradigmatic example, we compute the infrared (IR) spectrum of liquid water at ambient conditions from an MD trajectory obtained with a flexible water empirical force-field (q-TIP4P-f~\cite{habe+09jcp}). 
The dipole moments are either computed from the point charges of the force-field, or using a $(\lambda=1)$ MCoV model trained on the SCAN water dipoles dataset of Ref.~\citenum{Malosso2024}. The spectrum is then obtained as the Fourier transform of the dipole auto-correlation function,

\begin{align}
    S(\omega) = \frac{1}{3 \Omega k_B T} \int_{0}^{\infty} \langle \mathbf{J}(t) \cdot \mathbf{J}(0) \rangle e^{\mathrm{i} \omega t} \mathrm{d}t  
\end{align}
where $\mathbf{J}(t)=\dot{\boldsymbol{\mu}}$ is the dipole current. NVT MD at 300\,K is performed with \texttt{i-pi}~\cite{ceri+14cpc, litm+24jcp} and the IR spectrum is computed and filtered with \textsc{SporTran}~\cite{Ercole2022, Pegolo2025}. 
Indeed, ignoring the large blue-shift of the stretching peak---a known artifact of classical MD that can be corrected by approximate quantum dynamics techniques~\cite{ross+14jcp2}---\autoref{fig:water spectrum} shows that MCoV generally outperforms the point-charge spectrum, especially in the low-frequency region. The latter in fact overestimates the intensity of the bending peak and underestimates that of the stretching one. The MCoV model qualitatively aligns well with the experimental data taken from Ref.~\citenum{Bertie1996}, and is also able to reveal the hydrogen bond stretching peak around 200\,cm$^{-1}$, which is due to intermolecular charge fluctuations~\cite{Sharma2005} and cannot be observed from a purely geometrical charge model such as q-TIP4P-f.
We remark that this comparison can only be qualitative, as reaching quantitative agreement with with experiments would require performing MD directly at the SCAN-DFT level of theory, as exemplified in Ref.~\citenum{malosso2025arxiv}.

\subsection{Breaking the vector-only model: the case of CO\textsubscript{2}}\label{sec:results_CO2}

As discussed in \autoref{sec:limitations-and-correction}, the simple vector-based prediction of equivariants exhibits pathological behavior for symmetric structures and/or atomic environments. To show this, we train models with and without the $\lambda$-correction on a simple dataset of CO\textsubscript{2} molecules sampled from an MD trajectory at 300\,K performed with the PET-MAD potential~\cite{PET-MAD-2025}. The training targets need to be per-atom quantities with spherical components with $\lambda>1$. Examples of such quantities are Born effective charges $Z^\star_{I\alpha\beta}$ and the Raman tensor $\chi_{I,\alpha\beta\gamma}$, which are obtained from derivatives of the potential energy $U$ with respect to electric fields $\mathcal{E}_\alpha$ and the nuclear positions $r_{I\gamma}$:

\begin{align}
    Z^{\star}_{I\alpha\beta} &= - \frac{\partial^2 U}{\partial\mathcal{E}_\alpha \partial r_{I\beta}} \\
    \chi_{I,\alpha\beta\gamma} &= - \frac{\partial^3 U}{\partial\mathcal{E}_\alpha\partial\mathcal{E}_\beta\partial r_{I\gamma}}
\end{align}

\begin{figure*}[t!]
    \centering
    \includegraphics[width=.9\linewidth]{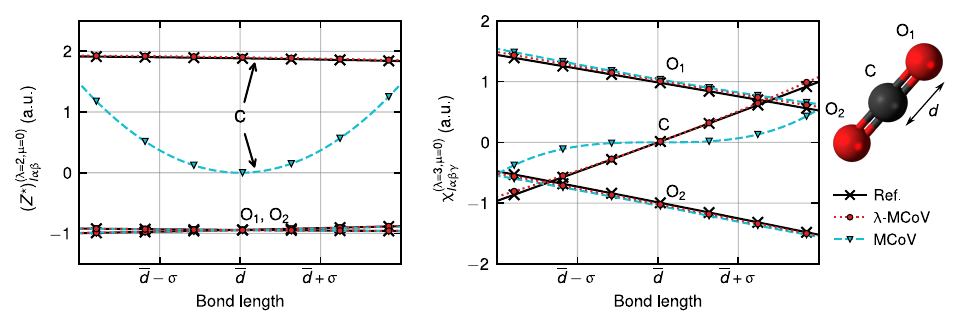}
    \caption{$(\lambda=2, \mu=0)$ component of the of the Born effective charges and $(\lambda=3, \mu=0)$ component of the Raman tensors of CO\textsubscript{2}.}
    \label{fig:co2}
\end{figure*}

We trained an MCoV and a $\lambda$-MCoV model targeting these quantities, which we computed with \textsc{PySCFAD} at the Hartree-Fock level, and we then evaluate it on a mock trajectory of CO\textsubscript{2} in a linear configuration aligned along the $z$ axis, with the carbon atom placed at the origin (0,0,0). The carbon and one of the oxygen atoms (O\textsubscript{2}) are kept fixed at a bond length $\overline{d}$, while the other oxygen atom (O\textsubscript{1}) is displaced such that its bond length varies from $\overline{d}-2\sigma$ to $\overline{d}+2\sigma$. Here, $\overline{d}$ is the equilibrium C–O bond length sampled along the MD trajectory, and $\sigma$ its standard deviation.

At the degeneracy point, the model trained without the $\lambda$-correction is constrained to predict zero for all tensorial properties with $\lambda \geq 1$, as well as for their derivatives up to order $(\lambda - 1)$, as analytically shown in the SI. This constraint does not apply to the actual target properties for the carbon atom. As shown in \autoref{fig:co2}, the Born effective charges do not vanish at the degeneracy, whereas those predicted by MCoV do, making them effectively impossible to learn. In contrast, the $\lambda$-MCoV model learns them with high accuracy.

A similar behavior is observed for the Raman tensor. In this case, the true property does pass through zero at the degeneracy, and both models reproduce this behavior. However, MCoV also predicts all derivatives to be zero, making the target function impossible to learn in the vicinity of the degeneracy.

\section{Discussion}

In this work, we have proposed a simple way to parametrize arbitrary tensors in three-dimensional space using scalar functions. 
This can be seen as the harmonic tensor generalization of the result that any equivariant vectorial function (with respect to the orthogonal group $O(3)$) lie in the space generated by its input~\cite{NEURIPS2021_f1b07759}.
Our main results show that such generalization holds also for the spherical components of a tensor of vectorial inputs, where the higher angular momentum are obtained by contracting the input vectors using only maximal CG contractions. 
We investigated the same generalization for pseudotensors, proving that the results apply if the CG contraction includes also one pseudovector, obtained from the standard cross product of vectors in the input. 

An application of interest regards the geometrical machine learning of quantum mechanical observables in atomic-scale systems, where symmetries and equivariance are fundamental. 
As the theoretical results exhibits an unpractical scaling with respect to the number of atoms in the local environments, a direct approach is not possible.

As a practical - if not rigorous - approximation, we propose a "$\lambda$-MCoV" model.
It uses a $\lambda=1$ spherical expansion to build a vectorial basis, and combine the vectors with maximal coupling to achieve the desired tensor order. 
Given that the basis becomes degenerate for high-symmetry configurations, this tensor is complemented with a spherical expansion of the same order, that we empirically found to be able to compensate for this deficiency. The architecture was then finalized by using a SOAP-BPNN model: capitalizing on the evaluation of the spherical expansion, the model evaluates the SOAP power spectrum, which is then fed to a BPNN to compute the necessary scalar functions.

We studied the performance of the model in several representative scenarios. First, we compared its accuracy against a simple $\lambda$-SOAP model on the multi-targets task of predicting dipole moments and polarizabilities on a subset of the QM7-X dataset.
In order to investigate higher angular momenta we trained a multi-target model that includes dipole, polarizability and hyperpolarizability for a subset of the QM7 dataset, while to investigate the performance for dynamical quantities we reproduced the water IR spectra. 
Finally, we trained a model for the Born effective charges and the Raman tensor, to study the performance on the model on per-atom quantities and investigate the effect of the correction on the case of highly symmetric configurations.

We believe that the simplicity and computational efficiency that can be achieved by predicting scalar functions, as opposed to using expensive equivariant neural networks, will make this method very valuable to applications where inference time is crucial, and require an exact description of the equivariant behavior of the properties of interest.

\section*{Data and software availability}

All software components used in this study are open-source and freely available. 
The Python package used to implement and train our ML models is \texttt{metatrain}, on GitHub at \url{https://github.com/metatensor/metatrain}.
The complete set of data and workflows required to reproduce all figures in this manuscript is provided in a Materials Cloud~\cite{tarl+20sd} repository~\cite{materialscloudrepo}. More information on how to reproduce the data are reported in the SI.

\begin{acknowledgments}
M.C. received funding from the European Research Council (ERC) under the European Union’s Horizon 2020 research
and innovation programme Grant No. 101001890-FIAMMA. F.B. and M.C. acknowledge support from the Swiss Plat-
form for Advanced Scientific Computing (PASC). All authors acknowledge funding from MARVEL National Centre of
Competence in Research (NCCR), funded by the Swiss National Science Foundation (SNSF, grant number 205602).

M.D. acknowledges and thank M. Langer and K. K. Huguenin-Dumittan for the insightful discussions. All the authors
thank S. Chong for comments on the data repositories, and J. W. Abbott for proof-reading the manuscript.
\end{acknowledgments}

\bibliography{others,biblio}
\onecolumngrid
\clearpage


\section*{Supplementary material}

- `https://pypi.org/project/pyscfad/`
- `https://github.com/fishjojo/pyscfad`

A tutorial on how to prepare the data and train a model with `metatrain` is available on the `atomistic-cookbook` at `https://atomistic-cookbook.org/examples/learn-tensors-with-mcov/learn-tensors-with-mcov.html`.

The complete set of data and workflows required to reproduce all figures in this manuscript is provided in a Materials Cloud [50] (DOI: `https://doi.org/10.24435/materialscloud:zq-a6`). The repository includes:

- `environment.yml`: Conda environment specification for all dependencies.
- `comparison_with_lambda_soap/`: Data and scripts to reproduce the results of Fig. 6.
- `train_multiple_equivariants_qm7/`: Data and scripts to reproduce the results of Fig. 7.
- `water_ir_spectrum/`: Data and scripts to reproduce the results of Fig. 8.
- `per_atom_equivariants_co2/`: Data and scripts to reproduce the results of Fig. 9.

## Appendices
## A Decomposing the $\lambda = 2$ contraction of two cross products

In this section we provide a proof for the identity

$$\big((a\,\widetilde{\otimes}\,b)_1\,\widetilde{\otimes}\,(c\,\widetilde{\otimes}\,d)_1\big)_{2\mu} = \frac{1}{2}\Big[(a\cdot c)\big(b\,\widetilde{\otimes}\,d\big)_{2\mu} + (b\cdot d)\big(a\,\widetilde{\otimes}\,c\big)_{2\mu} - (a\cdot d)\big(b\,\widetilde{\otimes}\,c\big)_{2\mu} - (b\cdot c)\big(a\,\widetilde{\otimes}\,d\big)_{2\mu}\Big]. \tag{A.1}$$

Recalling that $(a\,\widetilde{\otimes}\,b)_1$ is the representation, in terms of CG contractions, of $a \times b$, the formula states that the maximal contraction of cross products belongs to the space spanned by the CG contraction of proper vectors. We will prove this identity by means of the theory of re-coupling of angular momenta. The proof will be carried out in terms of the complex representation of spherical harmonics while the transformation in terms of real spherical harmonics will be performed only at the end. Given four arbitrary (complex-)harmonic tensors $T^{l_1}$, $U^{l_2}$, $V^{l_3}$ and $W^{l_4}$, belonging to four different angular momenta subspaces $l_1$, $l_2$, $l_3$ and $l_4$ we can perform a change of coupling scheme by means of the identity

$$\frac{1}{\sqrt{(2L_1+1)(2L_2+1)}}\big((T^{l_1}\,\widetilde{\otimes}\,U^{l_2})_{L_1}\,\widetilde{\otimes}\,(V^{l_3}\,\widetilde{\otimes}\,W^{l_4})_{L_2}\big)_{\lambda\mu}$$
$$= \sum_{L_1' L_2'} \sqrt{(2L_1'+1)(2L_2'+1)} \begin{Bmatrix} l_1 & l_3 & L_1' \\ l_2 & l_4 & L_2' \\ L_1 & L_2 & \lambda \end{Bmatrix} \big((T^{l_1}\,\widetilde{\otimes}\,V^{l_3})_{L_1'}\,\widetilde{\otimes}\,(U^{l_2}\,\widetilde{\otimes}\,W^{l_4})_{L_2'}\big)_{\lambda\mu},$$

where the coefficients in the curly brackets are the Wigner $9j$-symbols [1]. Importantly, these coefficients vanish unless each of the rows and columns satisfy the triangular inequality: For example, the first and second rows of the $9j$-symbol above select values of $L_1'$ and $L_2'$ such that $|l_1 - l_3| \leq L_1' \leq l_1 + l_3$ and $|l_2 - l_4| \leq L_2' \leq l_2 + l_4$ respectively. In particular the last column selects values such that the final contraction is compatible with the chosen $\lambda$ channel. Namely, the $9j$-symbol vanishes unless $|L_1' - L_2'| \leq \lambda \leq L_1' + L_2'$. Moreover, the $9j$-symbols is highly symmetric with respect to the exchange of rows and columns. Explicitly, they remain unchanged for any cyclic permutation of the columns and the rows, while they acquire a phase $(-1)^{l_1+l_2+l_3+l_4+L_1+L_2+L_1'+L_2'+\lambda}$ under an odd permutation.

Thus, using the $9j$-symbols and the relation above, it holds that:

$$\big((a\,\widetilde{\otimes}\,c)_0\,\widetilde{\otimes}\,(b\,\widetilde{\otimes}\,d)_2\big)_{2\mu} = \sqrt{5}\sum_{L_1 L_2}\sqrt{(2L_1+1)(2L_2+1)} \begin{Bmatrix} 1 & 1 & L_1 \\ 1 & 1 & L_2 \\ 0 & 2 & 2 \end{Bmatrix} \big((a\,\widetilde{\otimes}\,b)_{L_1}\,\widetilde{\otimes}\,(c\,\widetilde{\otimes}\,d)_{L_2}\big)_{2\mu}.$$

Now, we can apply the property $(a \,\widetilde{\otimes}\, b)_L = (-1)^L (b \,\widetilde{\otimes}\, a)_L$ (inherited by the same permutation property of the CG coefficients), and the identity

$$\begin{Bmatrix} 1 & 1 & L_1 \\ 1 & 1 & L_2 \\ 0 & 2 & 2 \end{Bmatrix} = (-1)^{L_1+L_2} \begin{Bmatrix} 1 & 1 & L_1 \\ 1 & 1 & L_2 \\ 2 & 0 & 2 \end{Bmatrix}.$$

to evaluate the following expression

$$\left((a\,\widetilde{\otimes}\,c)_0 \,\widetilde{\otimes}\, (b\,\widetilde{\otimes}\,d)_2\right)_{2\mu} + \left((a\,\widetilde{\otimes}\,c)_2 \,\widetilde{\otimes}\, (b\,\widetilde{\otimes}\,d)_0\right)_{2\mu} - \left((a\,\widetilde{\otimes}\,d)_0 \,\widetilde{\otimes}\, (b\,\widetilde{\otimes}\,c)_2\right)_{2\mu} - \left((a\,\widetilde{\otimes}\,d)_2 \,\widetilde{\otimes}\, (b\,\widetilde{\otimes}\,c)_0\right)_{2\mu}$$

$$= \sqrt{5} \sum_{L_1 L_2} \sqrt{(2L_1+1)(2L_2+1)} \begin{Bmatrix} 1 & 1 & L_1 \\ 1 & 1 & L_2 \\ 0 & 2 & 2 \end{Bmatrix} \left(1 - (-1)^{L_1} - (-1)^{L_2} + (-1)^{L_1+L_2}\right) \left((a\,\widetilde{\otimes}\,b)_{L_1} \,\widetilde{\otimes}\, (c\,\widetilde{\otimes}\,d)_{L_2}\right)_{2\mu}$$

$$= 12\sqrt{5} \begin{Bmatrix} 1 & 1 & 1 \\ 1 & 1 & 1 \\ 0 & 2 & 2 \end{Bmatrix} \left((a\,\widetilde{\otimes}\,b)_1 \,\widetilde{\otimes}\, (c\,\widetilde{\otimes}\,d)_1\right)_{2\mu} = -\frac{2}{\sqrt{3}} \left((a\,\widetilde{\otimes}\,b)_1 \,\widetilde{\otimes}\, (c\,\widetilde{\otimes}\,d)_1\right)_{2\mu}.$$

To obtain the last line, we noticed that the only values of $L_1$ and $L_2$ that are simultaneously compatible with the triangular inequalities of the $9j$-symbol and the expression in the round brackets are $L_1 = L_2 = 1$.

The above relation is already in a form close to Eq. (A.1). To conclude the proof we just need to recall that $(a\,\widetilde{\otimes}\,b)_0 = -(a \cdot b)/\sqrt{3}$, namely it is proportional to the standard scalar product between $a$ and $b$, and that $C^{\lambda\mu}_{00\lambda\mu} = 1$ for the CG coefficients. Therefore, we can write

$$\left((a\,\widetilde{\otimes}\,c)_0 \,\widetilde{\otimes}\, (b\,\widetilde{\otimes}\,d)_2\right)_{2\mu} = -\frac{1}{\sqrt{3}} (a \cdot c)(b\,\widetilde{\otimes}\,d)_{2\mu},$$

which leads directly to Eq. (A.1). Although we have derived the result for the complex harmonic representation, we notice that the real representation is obtained by multiplying both sides of the identity by the same transformation (unitary) matrix. Thus the identity also holds for the real harmonic representation, concluding the proof.

## B  Results for proper tensors

### B.1  Preliminaries on the spherical representation

This section is devoted to providing proofs of Proposition 1, Corollary 2 and Proposition 3. Before diving into the proofs we recall how to obtain the spherical components of an arbitrary Cartesian tensor $\mathscr{A}^{(\lambda)}$ of rank $\lambda$. Let us consider its expansion in terms of the canonical basis

$$\mathscr{A}^{(\lambda)} = \sum_{q_1 \ldots q_\lambda} A^{q_1 \ldots q_\lambda} \left(\hat{e}_{q_1} \otimes \ldots \otimes \hat{e}_{q_\lambda}\right), \tag{B.1}$$

where $A^{q_1 \ldots q_\lambda}$ are the Cartesian components of the tensor. We can now use the generalized CG coefficients (see Ref. 52, 34, 26, 27) defined as

$$C^{L_1 \ldots L_{\lambda-2}, \lambda\mu}_{1q_1 \ldots 1q_\lambda} := \sum_{M_2 \ldots M_{\lambda-1}} C^{L_1 M_1}_{1q_1 1q_2} C^{L_2 M_2}_{L_1 M_1 1q_3} \cdot \ldots \cdot C^{\lambda\mu}_{L_{\lambda-2} M_{\lambda-2} 1q_\lambda}, \tag{B.2}$$

namely, as progressive contractions of CG coefficients. From the orthogonality and completeness property of the CG coefficients [1],

$$\sum_{m_1 m_2} C^{lm*}_{l_1 m_1 l_2 m_2} C^{l'm'}_{l_1 m_1 l_2 m_2} = \delta_{ll'} \delta_{mm'} \qquad \text{and} \qquad \sum_{lm} C^{lm*}_{l_1 m_1 l_2 m_2} C^{lm}_{l_1 m'_1 l_2 m'_2} = \delta_{m_1 m'_1} \delta_{m_2 m'_2},$$

written in terms of Kronecker deltas, it is possible to derive an analogous property for the generalized CG coefficients

$$\sum_{LM} \sum_{L_1 \ldots L_{\lambda-2}} C^{L_1 \ldots L_{\lambda-2}, LM*}_{1q_1 \ldots 1q_\lambda} C^{L_1 \ldots L_{\lambda-2}, LM}_{1q'_1 \ldots 1q'_\lambda} = \delta_{q_1 q'_1} \cdot \ldots \cdot \delta_{q_\lambda q'_\lambda}, \tag{B.3}$$

and

$$\sum_{q_1 \ldots q_\lambda} C^{L_1 \ldots L_{\lambda-2}, LM*}_{1q_1 \ldots 1q_\lambda} C^{L'_1 \ldots L'_{\lambda-2}, L'M'}_{1q_1 \ldots 1q_\lambda} = \delta_{L_1 L'_1} \cdot \ldots \cdot \delta_{L_{\lambda-2} L'_{\lambda-2}} \delta_{LL'} \delta_{MM'}. \tag{B.4}$$

We remark that, in contrast with the CG coefficients defined for the standard spherical harmonics, which are defined as real coefficients, when dealing with a real representation of the spherical harmonics the CG coefficients are, generally, complex.

By means of these properties it is possible to move from the Cartesian representation to the spherical one. Indeed, it holds that

$$\begin{aligned}\mathscr{A}^{(\lambda)} &= \sum_{q_1...q_\lambda} A^{q_1...q_\lambda}(\hat{e}_{q_1} \otimes ... \otimes \hat{e}_{q_\lambda}) \\ &= \sum_{LM} \sum_{L_1...L_{\lambda-2}} \sum_{\substack{q_1...q_\lambda \\ q'_1...q'_\lambda}} A^{q_1...q_\lambda} C^{L_1...L_{\lambda-2},LM}_{1q_1...1q_\lambda} C^{L_1...L_{\lambda-2},LM}_{1q'_1...1q'_\lambda}(\hat{e}_{q'_1} \otimes ... \otimes \hat{e}_{q'_\lambda}) \\ &= \sum_{LM} \sum_{L_1...L_{\lambda-2}} A^{L_1...L_{\lambda-2},LM}_{\text{sph}} \left[\sum_{q_1...q_\lambda} C^{L_1...L_{\lambda-2},LM}_{1q_1...1q_\lambda}(\hat{e}_{q_1} \otimes ... \otimes \hat{e}_{q_\lambda})\right],\end{aligned} \quad \text{(B.5)}$$

where the last line defines the spherical components of the tensor as

$$A^{L_1...L_{\lambda-2},LM}_{\text{sph}} := \sum_{q_1...q_\lambda} C^{L_1...L_{\lambda-2},LM}_{1q_1...1q_\lambda} A^{q_1...q_\lambda}, \quad \text{(B.6)}$$

and the inverse relation as

$$A^{q_1...q_\lambda} = \sum_{LM} \sum_{L_1...L_{\lambda-2}} C^{L_1...L_{\lambda-2},LM}_{1q_1...1q_\lambda} A^{L_1...L_{\lambda-2},LM}_{\text{sph}}. \quad \text{(B.7)}$$

In general, the spherical components depend on the specific coupling scheme chosen but for the maximal case $L = \lambda$, which does is independent of the way in which the generalized CG coefficients are defined. Therefore, if $T_\lambda \in \mathbb{R}^{2\lambda+1}$ contains the harmonic components of a tensor belonging to the space of angular momentum $\lambda$, we can identify

$$A^{L_1...L_{\lambda-2},\lambda\mu}_{\text{sph}} \equiv T_{\lambda\mu} \quad \text{for} \quad L = \lambda, \quad \text{and} \quad A^{L_1...L_{\lambda-2},LM}_{\text{sph}} \equiv 0 \quad \text{if} \quad L \neq \lambda. \quad \text{(B.8)}$$

We can view these relations as a way to pad the tensor $T_\lambda$ in order to be able to reconstruct a Cartesian tensor of rank $\lambda$, setting to zero all the coupling path but the maximal one.

The tensor enclosed in the square brackets in the last line of Eq. (B.5) can be interpreted as a basis for the spherical representation. This can be seen from the relation

$$\left(\sum_{q_1...q_\lambda} C^{L_1...L_{\lambda-2},LM}_{1q_1...1q_\lambda}(\hat{e}_{q_1} \otimes ... \otimes \hat{e}_{q_\lambda})\right)^* \cdot \left(\sum_{q'_1...q'_\lambda} C^{L'_1...L'_{\lambda'-2},L'M'}_{1q'_1...1q'_\lambda}(\hat{e}_{q'_1} \otimes ... \otimes \hat{e}_{q'_\lambda})\right) = \delta_{L_1 L'_1} \cdot ... \cdot \delta_{L_{\lambda-2} L'_{\lambda-2}} \delta_{LL'} \delta_{MM'}, \quad \text{(B.9)}$$

obtained by means of the orthogonality of the canonical vectors and of the generalized CG coefficients of Eq. (B.4).

If we now take inspiration from the definition in Eq. (9) of the main text, and we define

$$\left(\hat{e} \,\widetilde{\otimes}\, ... \,\widetilde{\otimes}\, \hat{e}\right)_{\lambda\mu} := \sum_{q_1...q_\lambda} \widetilde{C}^{\lambda\mu}_{q_1...q_\lambda}(\hat{e} \otimes ... \otimes \hat{e})_{q_1...q_\lambda}, \quad \text{(B.10)}$$

by means of the maximal coupling, we can use this object to project the tensor $\mathscr{A}^{(\lambda)}$ onto the subspace spanned by its maximal components. Explicitly, we obtain the most useful property

$$T_\lambda = \left(\hat{e} \,\widetilde{\otimes}\, ... \,\widetilde{\otimes}\, \hat{e}\right)_\lambda \cdot \mathscr{A}^{(\lambda)}, \quad \text{(B.11)}$$

where we used the fact that the CG coefficients for the maximal coupling are real. The last relevant property is the effect of the spherical basis on a general maximal CG coupling of $\lambda$ vectors, namely

$$\left(\hat{e} \,\widetilde{\otimes}\, ... \,\widetilde{\otimes}\, \hat{e}\right)_{\lambda\mu} \cdot (a_1 \otimes ... \otimes a_\lambda) = \sum_{q_1...q_\lambda} \widetilde{C}^{\lambda\mu}_{q_1...q_\lambda}\left((a_1)_{q_1} \cdot ... \cdot (a_\lambda)_{q_\lambda}\right) = (a_1 \,\widetilde{\otimes}\, ... \,\widetilde{\otimes}\, a_\lambda)_{\lambda\mu}. \quad \text{(B.12)}$$

This relation shows that applying the maximal coupling has the effect of converting the external products into a maximal CG contraction.

## B.2 On the representation of proper tensors

**Proof of Proposition 1.** Given the set of input vectors $\{r_i\}_{i=1}^n \in (\mathbb{R}^3)^n$, we define $S = \text{span}(\{r_i\})$ and we call $S_\perp$ its orthogonal complement. Therefore, since $\mathbb{R}^3 = S \oplus S_\perp$, we can write

$$\hat{e}_q = \sum_{j=1}^n \alpha_q^j r_j + \sum_{k=1}^{n_{\text{ort}}} \beta_q^k p_k, \tag{B.13}$$

for some coefficients $\alpha_i^j$ and $\beta_i^k$, and where the vectors $p_k$ constitute a basis for the orthogonal complement $S_\perp$. Here we define $n_{\text{inp}} := \dim(S)$ and $n_{\text{ort}} := \dim(S_\perp)$ as the dimension of the space generated by the input vectors and of their orthogonal complement, respectively, and we have that $n_{\text{inp}} + n_{\text{ort}} = 3$. We notice here that this strategy closely follows the same proof outlined in Ref. 20, which will be recovered as a special case of this proof for $\lambda = 1$.

We can now plug the above relation in the general form of Eq. (B.1), apply the projection of Eq. (B.11) to recover the tensor $T_\lambda$ and exploit the effect of this projection of external products as indicated by Eq. (B.12) to obtain

$$\begin{aligned}T_\lambda(\{r_i\}) =& \sum_{j_1\ldots j_\lambda}\left(\sum_{q_1\ldots q_\lambda} A^{q_1\ldots q_\lambda}\alpha_{q_1}^{j_1}\cdot\ldots\cdot\alpha_{q_\lambda}^{j_\lambda}\right)\left(r_{j_1}\tilde{\otimes}\ldots\tilde{\otimes}r_{j_\lambda}\right)_\lambda \\ &+ \sum_{j_1\ldots j_{\lambda-1}k_\lambda}\left(\sum_{q_1\ldots q_\lambda} A^{q_1\ldots q_\lambda}\alpha_{q_1}^{j_1}\cdot\ldots\cdot\beta_{q_\lambda}^{j_\lambda}\right)\left(r_{j_1}\tilde{\otimes}\ldots\tilde{\otimes}p_{k_\lambda}\right)_\lambda \\ &\ldots \\ &+ \sum_{k_1\ldots k_\lambda}\left(\sum_{q_1\ldots q_\lambda} A^{q_1\ldots q_\lambda}\beta_{q_1}^{j_1}\cdot\ldots\cdot\beta_{q_\lambda}^{j_\lambda}\right)\left(p_{k_1}\tilde{\otimes}\ldots\tilde{\otimes}p_{k_\lambda}\right)_\lambda,\end{aligned} \tag{B.14}$$

where we took all possible combinations of vectors $r_j$ and vectors $p_k$. In the expression above, the rotation acts only on the CG contraction and not on the terms in round brackets, which behave as scalars. Further separating the terms with CG contractions containing an even number of vectors $p_k$ from the ones containing an odd number, leads to the partition

$$T_\lambda(\{r_i\}) = T_\lambda^{\text{even}}(\{r_i\}) + T_\lambda^{\text{odd}}(\{r_i\}). \tag{B.15}$$

Let us first show now that $T_\lambda^{\text{odd}} = 0$. This will be done in accord with the proof of Ref. 20, and is indeed a generalization of the argument therein: if only an even number of vectors generating $S_\perp$ can appear, and if $\lambda = 1$ (only one vector at a time is considered), then there cannot be any vector in $S_\perp$ at all. The proof goes at follows: taking an element $Q \in O(3)$, then the equivariance with respect of the inputs reads

$$Q^{(\lambda)} T_\lambda(\{r_i\}) = T_\lambda(\{Q^{(1)} r_i\}), \tag{B.16}$$

where $Q^{(\lambda)}$ is the appropriate representation of the element $Q$. We can now select a $Q$ such that its action is that $Q^{(1)} r_j = r_j$ and $Q^{(1)} p_k = -p_k$.[1] On the one hand, due to the equivariance condition, we have that

$$Q^{(\lambda)} T_\lambda(\{r_i\}) = T_\lambda(\{r_i\}) = T_\lambda^{\text{even}}(\{r_i\}) + T_\lambda^{\text{odd}}(\{r_i\}).$$

on the other hand, the addend containing an odd number of vectors $p_k$ change sign, namely

$$Q^{(\lambda)} T_\lambda(\{r_i\}) = T_\lambda^{\text{even}}(\{r_i\}) - T_\lambda^{\text{odd}}(\{r_i\}).$$

A comparison of the two expressions shows that, indeed, $T_\lambda^{\text{odd}} = 0$. While this was sufficient to prove the case $\lambda = 1$ in Ref. 20, here we have to go a step further to show that no vectors in $S_\perp$ can appear in Eq. (B.14). If the input vectors generate the whole space, $n_{\text{inp}} = 3$, there is nothing to prove as the orthogonal complement contains only the null vector. Let us investigate the remaining two cases, $n_{\text{ort}} = 1$ and $n_{\text{ort}} = 2$ separately.

**Case $\dim(S_\perp) = 1$.** In this case we have only one vector, $p$, generating $S_\perp$. We first notice that the maximal CG contractions are commutative, and therefore we can pair all the vectors $p$ together. Furthermore, since the maximal CG contraction does not depend on the specific coupling scheme, we can first couple the vectors $p$ together. Therefore, the CG contractions that appear in the expansion of $T_\lambda$ are all of the form

$$\left(r_{i_1}\tilde{\otimes}\ldots\tilde{\otimes}r_{i_s}\tilde{\otimes}(p\tilde{\otimes}p)_2\tilde{\otimes}\ldots\tilde{\otimes}(p\tilde{\otimes}p)_2\right)_{\lambda\mu}, \tag{B.17}$$

---

[1] The transformation $Q$ is realized in different ways depending on the dimensionality of the space $S$, $n_{\text{inp}}$. If $n_{\text{inp}} = 3$ this is simply the identity, as there are no vectors $p_k$. If $n_{\text{inp}} = 2$ the transformation is obtained by combining a global inversion and a rotation of $\pi$ radians around the axis defined by $S_\perp$. If $n_{\text{inp}} = 1$, the transformation is the rotation of $\pi$ around the direction of $S$.

for some integer $s$. Noticing that $\boldsymbol{p}$ is in the direction of the cross-product between two non-collinear vectors of $S$, and recalling that $(\boldsymbol{a} \,\widetilde{\otimes}\, \boldsymbol{b})_1 = -\mathrm{i}(\boldsymbol{a} \times \boldsymbol{b})/\sqrt{2}$, we can write

$$\boldsymbol{p} = \mathrm{i}\sum_{ij} \gamma_{ij}(\boldsymbol{r}_i \,\widetilde{\otimes}\, \boldsymbol{r}_j)_1, \qquad (B.18)$$

for some real coefficient $\gamma_{ij}$. Since the vector $\boldsymbol{p}$ can only appear an even number of times in the CG contractions, then the full expression is always real. In particular, each of the $(\boldsymbol{p} \,\widetilde{\otimes}\, \boldsymbol{p})_2$ terms can be written as

$$(\boldsymbol{p} \,\widetilde{\otimes}\, \boldsymbol{p})_2 = -\sum_{i_1 j_1 i_2 j_2} \gamma_{i_1 j_1} \gamma_{i_2 j_2} \Big((\boldsymbol{r}_{i_1} \,\widetilde{\otimes}\, \boldsymbol{r}_{j_1})_1 \,\widetilde{\otimes}\, (\boldsymbol{r}_{i_2} \,\widetilde{\otimes}\, \boldsymbol{r}_{j_2})_1\Big)_2. \qquad (B.19)$$

Using now the identity of Eq. (A.1) on the right-hand side proves that the CG contraction $(\boldsymbol{p} \,\widetilde{\otimes}\, \boldsymbol{p})_2$ entirely lies in the space generated by the pair-wise CG contractions $\{(\boldsymbol{r}_i \,\widetilde{\otimes}\, \boldsymbol{r}_j)_2\}_{i,j=1}^n$. Plugging this back into Eq. (B.17) proves that we can re-write the expansion of the tensor $T_\lambda$ only in terms of CG contractions of the form $(\boldsymbol{r}_{i_1} \,\widetilde{\otimes}\, \ldots \,\widetilde{\otimes}\, \boldsymbol{r}_{i_\lambda})_\lambda$, as desired.

**Case $\dim(S_\perp) = 2$.** In this extreme case, we have that the input vectors $\{\boldsymbol{r}_i\}_{i=1}^n$ are all collinear. Therefore, the orthogonal complement of $S_\perp$ is generated by two independent vectors $\boldsymbol{p}_1$ and $\boldsymbol{p}_2$. To manifest the symmetries of the tensors $T_\lambda$ in this configuration, we can align the $\hat{\boldsymbol{z}}$-axis of the frame of reference with the input space $S$. Furthermore we can make the identification $\hat{\boldsymbol{p}}_1 = \hat{\boldsymbol{x}}$ and $\hat{\boldsymbol{p}}_2 = \hat{\boldsymbol{y}}$.[2] With this choice, all the input is along the $\hat{\boldsymbol{z}}$ axis, requiring the tensor $T_\lambda$ to be cylindrically symmetric around the same axis. To show this, we will use the picture of complex spherical harmonics, as rotations and CG properties are much simplified in this formalism.[3] We will use this picture only in this paragraph and in the analogous ones regarding pseudotensors. Importantly, the $\mu = 0$ components of the tensor are the same in both real and complex representations. In this picture, a rotation around the $z$-axis is obtained by the simplified Wigner-$D$ matrix $D^{(\lambda)}_{\mu\mu'}(0,0,\phi) = e^{i\mu\phi}\delta_{\mu\mu'}$, where $\phi$ is the rotation angle. As the input remains unchanged under this rotation (it being oriented along the $\hat{\boldsymbol{z}}$ axis) the equivariance implies that

$$e^{i\mu\phi} T_{\lambda\mu}(\{\boldsymbol{r}_i\}) = T_{\lambda\mu}(\{\boldsymbol{r}_i\}), \qquad (B.20)$$

which shows that the all the components for $\mu \neq 0$ must vanish. We will prove that including terms with unit vectors $\hat{\boldsymbol{x}}$ and $\hat{\boldsymbol{y}}$ in Eq. (B.14) brings non-vanishing components for $\mu \neq 0$, which leads to a contradiction.

Firstly, we prove that the unit vectors $\hat{\boldsymbol{x}}$ and $\hat{\boldsymbol{y}}$ can appear only an even number of times in Eq. (B.14). This is straightforwardly achieved by using the same strategy that led to the constraints $T^{\text{odd}}_\lambda(\{\boldsymbol{r}_i\}) = 0$ in Eq. (B.15), with the two transformations $Q_x, Q_y \in O(3)$ such that they invert only one axis, namely

$$\begin{aligned} Q^{(1)}_x \boldsymbol{r}_i &= \boldsymbol{r}_i, & Q^{(1)}_x \hat{\boldsymbol{x}} &= -\hat{\boldsymbol{x}}, & \text{and} \quad Q^{(1)}_x \hat{\boldsymbol{y}} &= \hat{\boldsymbol{y}}, \\ Q^{(1)}_x \boldsymbol{r}_i &= \boldsymbol{r}_i, & Q^{(1)}_x \hat{\boldsymbol{x}} &= \hat{\boldsymbol{x}}, & \text{and} \quad Q^{(1)}_y \hat{\boldsymbol{y}} &= -\hat{\boldsymbol{y}}. \end{aligned} \qquad (B.21)$$

Thus, a general term in Eq. (B.14) has the form

$$\Big(\underbrace{\boldsymbol{r}_{i_1} \,\widetilde{\otimes}\, \ldots \,\widetilde{\otimes}\, \boldsymbol{r}_{i_{\lambda-2s-2t}}}_{\lambda-2s-2t \text{ terms}} \,\widetilde{\otimes}\, \underbrace{\hat{\boldsymbol{x}} \,\widetilde{\otimes}\, \ldots \,\widetilde{\otimes}\, \hat{\boldsymbol{x}}}_{2s \text{ terms}} \,\widetilde{\otimes}\, \underbrace{\hat{\boldsymbol{y}} \,\widetilde{\otimes}\, \ldots \,\widetilde{\otimes}\, \hat{\boldsymbol{y}}}_{2t \text{ terms}}\Big)_\lambda = \Big(\underbrace{\boldsymbol{r}_{i_1} \,\widetilde{\otimes}\, \ldots \,\widetilde{\otimes}\, \boldsymbol{r}_{i_{\lambda-l_1-l_2}}}_{\lambda-2s-2t \text{ terms}} \,\widetilde{\otimes}\, \hat{\boldsymbol{x}}^{\widetilde{\otimes}(2s)} \,\widetilde{\otimes}\, (\hat{\boldsymbol{y}} \,\widetilde{\otimes}\, \hat{\boldsymbol{y}})_2^{\widetilde{\otimes}t}\Big)_\lambda, \qquad (B.22)$$

where we used the shorthand notations for the self contraction $\hat{\boldsymbol{x}}^{\widetilde{\otimes}s} := (\hat{\boldsymbol{x}} \,\widetilde{\otimes}\, \ldots \,\widetilde{\otimes}\, \hat{\boldsymbol{x}})_s$. In the last term, we paired the unit vectors $\hat{\boldsymbol{y}}$ in a way similar to what (B.17), and which allows to follow the same reasoning: each of these vectors can be written in terms of the vectors in the orthogonal plane only. Explicitly, as we can write $\hat{\boldsymbol{y}} = \mathrm{i}\sum_i \gamma_i(\boldsymbol{r}_i \,\widetilde{\otimes}\, \hat{\boldsymbol{x}})_1$ for some coefficients $\gamma_i$, then using the identity of Eq. (A.1) we obtain

$$(\hat{\boldsymbol{y}} \,\widetilde{\otimes}\, \hat{\boldsymbol{y}})_2 = -\frac{1}{2}\sum_{ij} \gamma_i \gamma_j \big[(\boldsymbol{r}_i \cdot \boldsymbol{r}_j)(\hat{\boldsymbol{x}} \,\widetilde{\otimes}\, \hat{\boldsymbol{x}})_2 + (\boldsymbol{r}_i \,\widetilde{\otimes}\, \boldsymbol{r}_j)_2\big], \qquad (B.23)$$

where we also exploited the orthogonality between $\hat{\boldsymbol{x}}$ and every vector in the input. Therefore, the general term appearing in Eq. (B.14) has the form

$$\Big(\underbrace{\boldsymbol{r}_{i_1} \,\widetilde{\otimes}\, \ldots \,\widetilde{\otimes}\, \boldsymbol{r}_{i_{\lambda-2s}}}_{\lambda-2s \text{ terms}} \,\widetilde{\otimes}\, \hat{\boldsymbol{x}}^{\widetilde{\otimes}(2s)}\Big)_\lambda \propto \sigma_{i_1} r_{i_1} \cdots \sigma_{i_{\lambda-2s}} r_{i_{\lambda-2s}} (Y_{\lambda-2s}(\hat{\boldsymbol{z}}) \,\widetilde{\otimes}\, Y_{2s}(\hat{\boldsymbol{x}}))_\lambda. \qquad (B.24)$$

---

[2]This choice is not restrictive as the frame of reference can be rotated to have the desired orientation and we can choose any pairs of non-collinear vectors in $S_\perp$ to construct a basis.

[3]Again, this is not restrictive as the two representations are linked by a linear (unitary) transformation [53], and all the operations in our work are linear.

Here we used the fact that the self contraction of a vector is proportional to the solid spherical harmonics, $a^{\widetilde{\otimes} s} \propto |a| Y_s(\hat{a})$, by some unessential proportionality factor. We also used the alignment of all the vectors in $S$ with the $\hat{z}$ axis, with the coefficients $\sigma_i$ being just $\pm 1$, depending on the position of the vector $r_i$. The final step is to explicitly evaluate

$$(Y_{\lambda-2s}(\hat{z}) \widetilde{\otimes} Y_{2s}(\hat{x}))_{\lambda\mu} = \sum_{m_1 m_2} C^{\lambda\mu}_{(\lambda-2s)m_1(2s)m_2} Y_{(\lambda-2s)m_1}(\hat{z}) Y_{(2s)m_2}(\hat{x}) = \sqrt{\frac{2l+1}{4\pi}} C^{\lambda\mu}_{(\lambda-2s)0(2s)\mu} Y_{(2s)\mu}(\hat{x}), \quad (B.25)$$

where we used the identity $Y_{(\lambda-2s)m_1}(\hat{z}) = \delta_{m_1 0} \sqrt{(2l+1)/(4\pi)}$, and the selection rule of the CG coefficients, $C^{lm}_{l_1 m_1 l_2 m_2}$ that require $m_1 + m_2 = m$. We observe that for $\mu > 2s$ all the components are zero, because of the same selection rules. Instead, for $\mu = 2s$, both the CG coefficient and the spherical harmonic are not zero. Thus, we obtain the proof: not only we have a non-zero component for $\mu \neq 0$, but varying $s$ brings always different components, not allowing for any simplification. Since we require that only the $\mu = 0$ component is not zero, we can accept only the $s = 0$ case in Eq. (B.24), namely the tensor lies in the space spanned by all the possible $(r_{i_1} \widetilde{\otimes} \ldots \widetilde{\otimes} r_{i_\lambda})_\lambda$. □

**Proof of Corollary 2.** The first half of this Corollary has the exact same proof as the analogous one from Ref. 20: given that the tensor $T_\lambda$ lies on the space generated by $(r_{i_1} \widetilde{\otimes} \ldots \widetilde{\otimes} r_{i_\lambda})_\lambda$, then we can write

$$T_\lambda(\{r_i\}) = \sum_{i_1 \ldots i_\lambda} a_{i_1 \ldots i_\lambda} (r_{i_1} \widetilde{\otimes} \ldots \widetilde{\otimes} r_{i_\lambda})_\lambda, \quad (B.26)$$

for some scalar $a_{i_1 \ldots i_\lambda}$. Performing the identification $f_{i_1 \ldots i_\lambda}(\{r_i\}) := a_{i_1 \ldots i_\lambda}$ for every tuple of vectors $\{r_i\}_{i=1}^n$ belonging to the same $O(3)$-orbit allows to write the above relation in terms of scalar functions. Moreover, due to the fact that the CG contractions are commutative, we can factorize functions that differ by swap of the indices and by re-defining the scalar functions we obtain

$$T_\lambda(\{r_i\}) = \sum_{i_1 \leq i_2 \leq \ldots \leq i_\lambda} f_{i_1 \ldots i_\lambda}(\{r_i\}) (r_{i_1} \widetilde{\otimes} \ldots \widetilde{\otimes} r_{i_\lambda})_\lambda, \quad (B.27)$$

for some functions $f_{i_1 \ldots i_\lambda}$, which are symmetric under indices permutations. This concludes the first half of the proof. The second half is obtained by a simple rearrangement of the functions $f_{i_1 \ldots i_\lambda}$. This means that these functions are uniquely characterized by how many times the same vector appears in the CG contraction, and not by the actual position of the index. Thus we can introduce the new definition

$$\Phi_{l_1 \ldots l_n}(\{r_i\}) := f_{\underbrace{1 \ldots 1}_{l_1 \text{ times}} \underbrace{2 \ldots 2}_{l_2 \text{ times}} \ldots \underbrace{n \ldots n}_{l_n \text{ times}}}(\{r_i\}), \quad \text{with} \quad l_1 + \ldots + l_n = \lambda,$$

which counts the number of times that the same vector appears in the summation. Importantly, the summation $l_1 + \ldots + l_n$ must be constrained to be equal to $\lambda$ since this is the total number of indices (indeed, $l_i$ is 0 if the corresponding index does not appear in the function $\beta$). In this way, the previous expression becomes

$$T_\lambda(\{r_i\}) = \sum_{\substack{l_1, \ldots, l_n = 0 \\ l_1 + \ldots + l_n = \lambda}}^{\lambda} \Phi_{l_1 \ldots l_n}(\{r_i\}) \big( \underbrace{r_1 \widetilde{\otimes} \ldots \widetilde{\otimes} r_1}_{l_1} \widetilde{\otimes} \underbrace{r_2 \widetilde{\otimes} \ldots \widetilde{\otimes} r_2}_{l_2} \widetilde{\otimes} \ldots \widetilde{\otimes} \underbrace{r_n \widetilde{\otimes} \ldots \widetilde{\otimes} r_n}_{l_n} \big)_\lambda. \quad (B.28)$$

Note that the function $\Phi_{l_1 \ldots l_n}$ is, in general, not symmetric under the swap of indices. Moreover we remark that the CG contraction is still maximal because of the constraint on the sum $l_1 + \ldots + l_n = \lambda$. Employing the definition of Eq. (10) in the main text concludes the proof of the Corollary. □

**Proof of Proposition 3.** This proof takes inspiration from the analogous one of [20], while expanding on it to be able to capture the tensorial nature of $T_\lambda$, since comparing scalar products in not as trivial when dealing with higher rank objects. Indeed, we will proceed here by induction on $\lambda$, starting from the simpler case $\lambda = 2$.

**Base Case ($\lambda = 2$).** For this case, starting from the components $T_2$, let us assume we did a padding similar to the one introduced at the beginning of this section and ended up with the Cartesian vector $\mathscr{A}$ with components $A^{q_1 q_2}$.

We remark that the transformation from Cartesian to spherical components and its inverse is given by a linear transformation, meaning that $A^{q_1 q_2}(\{r_i\})$ is still a polynomial function of the inputs. Another operation that does not change its polynomial nature is the contraction with an arbitrary unit vector $\hat{s}$, given by

$$\sum_{q_2} A^{q_1 q_2}(\{r_i\})(\hat{s})_{q_2} = \sum_{i=1}^n f_i(r_1, \ldots, r_n, \hat{s})(r_i)_{q_1} + f_s(r_1, \ldots, r_n)(\hat{s})_{q_1}. \quad (B.29)$$

Here we used the results of Ref. 20 since the left-hand side of the equation is a polynomial vectorial function: therefore, not only it lies in the space generated by the input (which in this case contains also the unit vector $\hat{s}$), but it does so by means of scalar polynomial functions, here indicated by $f_i$ and $f_s$. Since the left-hand side is an homogeneous polynomial of degree 1 in $\hat{s}$, the function $f_s$ cannot depend on it. We can now use the fundamental theorem of invariant theory (see Ref. 20) and expand the scalar functions $f_i$ in products of powers of the scalar products $(r_i \cdot r_j)$ and $(r_i \cdot \hat{s})$. Again, due the fact that the expression is homogeneous of degree 1 in $\hat{s}$, all the resulting terms must contain one, and only one, scalar product containing $\hat{s}$. Factorizing all the rest we obtain the expression

$$\sum_{q_2} A^{q_1 q_2}(\{r_i\})(\hat{s})_{q_2} = \sum_{ij} f_{ij}(r_1,\ldots,r_n)(r_j \cdot \hat{s})(r_i)_{q_1} + f_s(r_1,\ldots,r_n)(\hat{s})_{q_1}, \tag{B.30}$$

for some scalar and still polynomial functions $f_{ij}$. Now, we can recall that we can rewrite the unit vector $\hat{s}$ in terms of spherical harmonics as $(\hat{s})_q = \sqrt{4\pi/3}\, Y_1^q(\hat{s})$. Therefore, given the orthogonality of the (real) spherical harmonics,

$$\int_\Omega d\hat{s}\, Y_l^m(\hat{s}) Y_{l'}^{m'}(\hat{s}) = \delta_{ll'}\delta_{mm'},$$

we have that

$$\sqrt{\frac{3}{4\pi}} \int_\Omega d\hat{s}\, Y_1^{q_2}(\hat{s}) \sum_{q_2'} A^{q_1 q_2'}(\{r_i\})(\hat{s})_{q_2'} = A^{q_1 q_2}, \tag{B.31}$$

where the integral is performed over the surface of the unit sphere. In words, by means of this integral we are able to select a specific component of the Cartesian tensor. At the same time, if we perform the integral against the expansion above, we obtain

$$\sqrt{\frac{3}{4\pi}} \int_\Omega d\hat{s}\, Y_1^{q_2}(\hat{s}) \sum_{q_2'} A^{q_1 q_2'}(\{r_i\})(\hat{s})_{q_2'} = \sum_{ij} f_{ij}(r_1,\ldots,r_n)(r_i)_{q_1}(r_j)_{q_2} + f_s(r_1,\ldots,r_n)\delta_{q_1 q_2}. \tag{B.32}$$

Performing now the maximal CG contraction to the $\lambda = 2$ channel to extract the maximal spherical representation, and recalling that this operation extracts the components $T_\lambda$ from $\mathscr{A}$, we obtain

$$T_{2\mu}(\{r_i\}) = \sum_{q_1 q_2} C^{2\mu}_{1q_1 1q_2} A^{q_1 q_2} = \sum_{ij} f_{ij}(r_1,\ldots,r_n)(r_i \widetilde{\otimes} r_j)_{2\mu}, \tag{B.33}$$

where the sum containing the Kronecker delta was identically zero (the identity matrix does not possess a $\lambda = 2$ component). Being the functions $f_{ij}$ polynomials, this step proves the base case.

**Inductive step.** Now we assume that the statement holds for an harmonic tensor of angular momentum $(\lambda - 1)$ and we prove for $\lambda$. Again, we begin with the Cartesian (padded) representation $\mathscr{A}^{(\lambda)}$ for the tensor $T_\lambda$ and first we perform a contraction of the last component with an arbitrary unit vector $\hat{s}$, and then a maximal CG contraction. In doing so, we obtain

$$\sum_{q_1 \ldots q_{\lambda-1}} \mathscr{C}^{(\lambda-1)\mu'}_{q_1 \ldots q_{\lambda-1}} \sum_{q_\lambda} A^{q_1 \ldots q_\lambda}(\{\mathbf{r}_i\})(\hat{s})_{q_\lambda}$$
$$= \sum_{i_1 \ldots i_{\lambda-1}} h_{i_1 \ldots i_{\lambda-1}}(\{\mathbf{r}_i\},\hat{s})(r_{i_1} \widetilde{\otimes} \ldots \widetilde{\otimes} r_{i_{\lambda-1}})_{(\lambda-1)\mu'} + \sum_{i_1 \ldots i_{\lambda-2}} g_{i_1 \ldots i_{\lambda-2}}(\{\mathbf{r}_i\})(r_{i_1} \widetilde{\otimes} \ldots \widetilde{\otimes} r_{i_{\lambda-2}} \widetilde{\otimes} \hat{s})_{(\lambda-1)\mu'}, \tag{B.34}$$

where, in the second line, we used the inductive hypothesis since the first line is an (equivariant) polynomial harmonic tensor of angular momentum $(\lambda - 1)$. Here, $h$ and $g$ are polynomial scalar functions in their arguments. In particular, the presence of $\hat{s}$ in the second addend stems from the fact that, again, the expression is of degree 1 in the unit vector $\hat{s}$. For the same reason, and using again the fundamental theorem of invariant theory on the first function, we can write

$$h_{i_1 \ldots i_{\lambda-1}}(\{\mathbf{r}_i\},\hat{s}) = \sum_{i_\lambda} f_{i_1 \ldots i_\lambda}(\{\mathbf{r}_i\})(r_{i_\lambda} \cdot \hat{s}), \tag{B.35}$$

for some polynomial functions $f$ obtained by preserving the homogeneity of the expression with respect to $\hat{s}$ and by factoring out all the terms that do not contain it. From this point, we follow the same recipe outlined for the base case: We first integrate against the spherical harmonic $Y_1^{q_\lambda}(\hat{s})$ to remove the dependence on the arbitrary unit vector and select the $q_\lambda$ component. As an additional step we also contract with the CG coefficients, in order to recover the original maximal coupling. Explicitly we have the following chain of identities

$$\sum_{q_\lambda \mu'} C^{\lambda\mu}_{(\lambda-1)\mu'1q_\lambda} \sqrt{\frac{3}{4\pi}} \int_\Omega d\hat{s}\, Y_1^{q_\lambda}(\hat{s}) \left[ \sum_{q_1...q_{\lambda-1}} \mathscr{C}^{(\lambda-1)\mu'}_{q_1...q_{\lambda-1}} \sum_{q'_\lambda} A^{q_1...q_{\lambda-1}q'_\lambda}(\hat{s})_{q'_\lambda} \right] = \tag{B.36}$$

$$= \sum_{q_1...q_{\lambda-1}} \sum_{q_\lambda} \left[ \sum_{\mu'} C^{\lambda\mu}_{(\lambda-1)\mu'1q_\lambda} \mathscr{C}^{(\lambda-1)\mu'}_{q_1...q_{\lambda-1}} \right] A^{q_1...q_{\lambda-1}q_\lambda} = \sum_{q_1...q_\lambda} \mathscr{C}^{\lambda\mu}_{q_1...q_{\lambda-1}q_\lambda} A^{q_1...q_{\lambda-1}q_\lambda} = T_{\lambda\mu},$$

where we used the fact that the contraction in square brackets is the maximal coupling, and that the maximal coupling of the Cartesian components is our target tensor $T_\lambda$. We can now apply the same transformation to the right-hand side of Eq. (B.34). In particular, we notice that the final CG tensor product in the second addend becomes

$$\sqrt{\frac{3}{4\pi}} \int_\Omega d\hat{s}\, Y_1^{q_\lambda}(\hat{s})(r_{i_1} \widetilde{\otimes} ... \widetilde{\otimes} r_{i_{\lambda-2}} \widetilde{\otimes} \hat{s})_{(\lambda-1)\mu'} = \sum_{q_\lambda \mu''} C^{(\lambda-1)\mu'}_{(\lambda-2)\mu''1q_\lambda}(r_{i_1} \widetilde{\otimes} ... \widetilde{\otimes} r_{i_{\lambda-2}})_{(\lambda-2)\mu''}. \tag{B.37}$$

Since the triangular property of the CG coefficients implies that $(\lambda-2)$ vectors cannot be contracted in an equivariant-preserving way such that the result has angular momentum $\lambda$, this term must be dropped. On the other hand, applying the integral to the function $h_{i_1...i_{\lambda-1}}$ causes the selection of the $q_\lambda$ component of $r_{i_\lambda}$, as can be seen from the right-hand side of Eq. (B.35). Thus, performing the last CG contraction and putting everything together, we obtain

$$T_\lambda = \sum_{i_1...i_{\lambda-1}} f_{i_1...i_\lambda}(\{r_i\})(r_{i_1} \widetilde{\otimes} ... \widetilde{\otimes} r_{i_\lambda})_\lambda, \tag{B.38}$$

with $f_{i_1...i_\lambda}(\{r_i\})$ polynomial scalar functions, thus satisfying the inductive step and concluding the proof. □

## C  Results for pseudotensors

In this section, we will provide the proofs of Proposition 4, Corollary 5 and Proposition 6. While all these construction will be quite similar, some care is necessary when dealing with pseudotensors, given their different behavior under inversion.

### C.1  Preliminaries on the spherical representation for pseudotensors

Here we will use a construction analogous to what already shown in Sec. B.1 but applied to a pseudotensor. In this case, in order to accommodate a pseudotensor of angular momentum $\lambda$ into a Cartesian tensor, we require a rank $(\lambda+1)$. This is a direct consequence of the fact that a pseudovector can be represented as a bivector (for example, the cross product). Thus, from a general Cartesian tensor of rank $(\lambda+1)$

$$\mathscr{B}^{(\lambda+1)} = \sum_{q_0 q_1...q_\lambda} B^{q_0 q_1...q_\lambda}(\hat{e}_{q_0} \otimes \hat{e}_{q_1} \otimes ... \otimes \hat{e}_{q_\lambda}), \tag{C.1}$$

we can again extract the spherical components as

$$B^{L_0 L_1...L_{\lambda-2},LM}_{\mathrm{sph}} := \sum_{q_0...q_\lambda} C^{L_0 L_1...L_{\lambda-2},LM}_{1q_0...1q_\lambda} B^{q_0...q_\lambda}, \tag{C.2}$$

by means of the generalized CG coefficients defined in Eq. (B.2). We remark that the Cartesian components and the intermediate channels are here enumerated from 0 (for example we are enumerating from $L_0$ instead of $L_1$), in order to mirror more effectively the formalism of the previous section. Now, having a pseudotensor $\Theta_\lambda$, we can perform the identification

$$\begin{cases} B^{L_0...L_{\lambda-2},LM}_{\mathrm{sph}} \equiv i\Theta_{\lambda\mu}, & \text{for} \quad L_0 = 1, \quad L = \lambda, \quad \text{and} \quad M = \mu, \\ B^{L_0...L_{\lambda-2},LM}_{\mathrm{sph}} \equiv 0 & \text{otherwise.} \end{cases} \tag{C.3}$$

This identification is again unique as, by fixing $L_0 = 1$, it creates a setting which is analogous to the coupling of $\lambda$ vectors into the space of angular momentum $\lambda$, with the only difference that one of the vectors is a pseudovector. Therefore, if $L = \lambda$, we are again coupling $\lambda$ vectors (of which one is pseudo) into the space of maximal angular momentum, and thus the value of all the other intermediate channels must be the highest possible. In particular, following the path of maximal coupling, we have $L_1 = 2$, $L_3 = 2$, and up to $L_{\lambda-2} = (\lambda-1)$. Formally, the presence of the imaginary unit is necessary as both the Cartesian components $B^{q_0 q_1...q_\lambda}$ and the components $\Theta_{\lambda\mu}$ are chosen to be real, while the

CG contraction under the constraints above is an imaginary number. This can be easily seen by seen the action of the chosen CG contraction on the canonical basis. By following the same procedure of Eq. (B.5) on the path for that has non-vanishing component with respect to Eq. (C.3) we get

$$\sum_{qq_0q_1\ldots q_\lambda} C^{1q}_{1q_01q_1} \mathscr{C}^{\lambda\mu}_{qq_2\ldots q_\lambda} (\hat{e}_{q_0} \otimes \hat{e}_{q_1} \otimes \ldots \otimes \hat{e}_{q_\lambda}) = \sum_{qq_2\ldots q_\lambda} \mathscr{C}^{\lambda\mu}_{qq_2\ldots q_\lambda} \left( \left( \sum_{q_0q_1} C^{1q}_{1q_01q_1} \hat{e}_{q_0} \otimes \hat{e}_{q_1} \right) \otimes \hat{e}_{q_2} \otimes \ldots \otimes \hat{e}_{q_\lambda} \right), \tag{C.4}$$

where we exploited the fact that, but for the first coupling, the path followed is the one of maximal coupling. We can now recognize the innermost CG coupling as being $(\hat{e}_{q_0} \widetilde{\otimes} \hat{e}_{q_1})_{1q}$, and recall that this is related to the cross product by means of $(\hat{e}_{q_0} \widetilde{\otimes} \hat{e}_{q_1})_1 = -\mathrm{i}(e_{q_0} \times e_{q_1})/\sqrt{2}$. Since the contraction produced by the maximal coupling produces is real we have that, due to the imaginary unity, the spherical basis is imaginary and therefore also the relative spherical components must be as well. This relation between pseudovectors and the imaginary unity will be used extensively in the following[4] To further exploit the analogy with the formalism from the proper tensors, we can define canonical pseudovectors as

$$\hat{\boldsymbol{\omega}}_q := \sum_{q_0q_1} C^{1q}_{1q_01q_1} (\hat{e}_{q_0} \widetilde{\otimes} \hat{e}_{q_1})_{1q}, \tag{C.5}$$

which allows to easily recover the components of the pseudotensor by means of the projection (compare with Eq. (B.11))

$$\boldsymbol{\Theta}_\lambda = -\mathrm{i}(\hat{\boldsymbol{\omega}}_q \widetilde{\otimes} \hat{e}_{q_2} \widetilde{\otimes} \ldots \widetilde{\otimes} \hat{e}_{q_\lambda})^*_\lambda \cdot \mathscr{B}^{(\lambda+1)} = \mathrm{i}(\hat{\boldsymbol{\omega}}_q \widetilde{\otimes} \hat{e}_{q_2} \widetilde{\otimes} \ldots \widetilde{\otimes} \hat{e}_{q_\lambda})_\lambda \cdot \mathscr{B}^{(\lambda+1)}. \tag{C.6}$$

Here, the last line exploited the fact that the term in curly brackets is purely imaginary. Another important property is the fact that this "basis" has a straightforward effect on external products

$$(\hat{\boldsymbol{\omega}}_q \widetilde{\otimes} \hat{e}_{q_2} \widetilde{\otimes} \ldots \widetilde{\otimes} \hat{e}_{q_\lambda})_\lambda \cdot (a_0 \otimes a_1 \otimes a_2 \ldots \otimes a_\lambda) = \left((a_0 \widetilde{\otimes} a_1)_1 \widetilde{\otimes} a_2 \ldots \widetilde{\otimes} a_\lambda\right)_\lambda, \tag{C.7}$$

namely it convert the external products into a maximal CG contraction but for the first one, which is projected into a pseudovector (the cross product). These properties can now be used in a way that is completely analogous to what done for the proper tensors, as will be shown in the next section.

## D  On the representation of pseudotensors

**Proofs of Proposition 4 and Corollary 5.** As per title, this section is devoted to the proofs of Proposition 4 and Corollary 2. However, being these proofs almost identical to the analogous ones for proper tensors, we will only give a general outline and explicitly investigate only the case when the proofs diverge.
Retracing the Proof for Proposition 4, we first use Eq. B.13 to write the canonical basis in terms of the space $S$ and its orthogonal complement $S_\perp$. Then we can write the pseudotensor $\boldsymbol{\Theta}_\lambda$ in a way analogous to Eq. (B.14), with the only exception that also the pseudovectors $\hat{\boldsymbol{\omega}}_q$ appear in the expansion. Furthermore, by means of the same transformation $Q$, we can prove that only an even number of vectors in the orthogonal complement $S_\perp$ can appear in the expansion of $\boldsymbol{\Theta}_\lambda$. The generalization of the cases $\dim(S) = 3$ is straightforward in this case as well. However, due to the presence of the pseudovectors, a slightly different scenario appears when $\dim(S) = 2$. In this case, only one vector $\boldsymbol{p}$ is required to describe $S_\perp$. Since this vector can also appear in forming a pseudovectorial quantity, we have terms of the form

$$\left((\boldsymbol{p} \widetilde{\otimes} \boldsymbol{r}_{i_1})_1 \widetilde{\otimes} \boldsymbol{p} \widetilde{\otimes} \boldsymbol{r}_{i_2} \widetilde{\otimes} \ldots \widetilde{\otimes} \boldsymbol{r}_{i_{\lambda-1}}\right)_\lambda = \left(((\boldsymbol{p} \widetilde{\otimes} \boldsymbol{r}_{i_1})_1 \widetilde{\otimes} \boldsymbol{p})_2 \widetilde{\otimes} \boldsymbol{r}_{i_2} \widetilde{\otimes} \ldots \widetilde{\otimes} \boldsymbol{r}_{i_{\lambda-1}}\right)_\lambda, \tag{D.1}$$

which were not present in the proof for proper tensors. The equivalence with the right-hand side, which holds since we can freely choose the coupling scheme in the maximal coupling, will be exploited shortly. We remark that this is the only possibility, as $(\boldsymbol{p} \widetilde{\otimes} \boldsymbol{p})_1 = 0$ and every pairs of vectors $(\boldsymbol{p} \widetilde{\otimes} \boldsymbol{p})_2$ outside can be written in terms of $(\boldsymbol{r}_i \widetilde{\otimes} \boldsymbol{r}_j)_2$ by means of Eq. (A.1), namely we can consider only one vector $\boldsymbol{p}$ outside the pseudovector. However, also this terms can be easily manipulated to remove the explicit dependence from $\boldsymbol{p}$: exploiting Eq. B.18 we obtain

$$\begin{aligned}
\left((\boldsymbol{p} \widetilde{\otimes} \boldsymbol{r}_{i_1})_1 \widetilde{\otimes} \boldsymbol{p}\right)_2 &= \mathrm{i} \sum_{j_1 j_2} \gamma_{j_1 j_2} \left((\boldsymbol{p} \widetilde{\otimes} \boldsymbol{r}_{i_1})_1 \widetilde{\otimes} (\boldsymbol{r}_{j_1} \widetilde{\otimes} \boldsymbol{r}_{j_2})_1\right)_2 \\
&= \frac{\mathrm{i}}{2} \sum_{j_1 j_2} \gamma_{j_1 j_2} \left[(\boldsymbol{r}_{i_1} \cdot \boldsymbol{r}_{j_2})(\boldsymbol{p} \widetilde{\otimes} \boldsymbol{r}_{j_2})_2 - (\boldsymbol{r}_{i_1} \cdot \boldsymbol{r}_{j_1})(\boldsymbol{p} \widetilde{\otimes} \boldsymbol{r}_{j_1})_2\right] \\
&= -\frac{1}{2} \sum_{j_1 j_2 j_3 j_4} \gamma_{j_1 j_2} \gamma_{j_3 j_4} \left[(\boldsymbol{r}_{i_1} \cdot \boldsymbol{r}_{j_2})\left((\boldsymbol{r}_{j_3} \widetilde{\otimes} \boldsymbol{r}_{j_4})_1 \widetilde{\otimes} \boldsymbol{r}_{j_2}\right)_2 - (\boldsymbol{r}_{i_1} \cdot \boldsymbol{r}_{j_1})\left((\boldsymbol{r}_{j_3} \widetilde{\otimes} \boldsymbol{r}_{j_4})_1 \widetilde{\otimes} \boldsymbol{r}_{j_1}\right)_2\right],
\end{aligned} \tag{D.2}$$

---

[4]Technically, this is due to the fact that, in $\mathbb{R}^3$, vectors and pseudovectors (or equivalently bivectors) are one the Hodge dual of the other: For example, in the language of Geometric Algebra, they are related by the pseudo scalar $I$, for which it holds that $I^2 = -1$.

where, in the second line, we used Eq. (A.1) and the fact that $p$ is orthogonal to any vector in the input. Instead in the last line we used again Eq. (B.18) on the remaining vectors $p$. Plugging the last line into the equation above is the last step for the proof of the case $\dim(S) = 2$.

Also the case for $\dim(S) = 1$, in which the input vectors are all aligned, is easily proven as no equivariant pseudotensor can exist. To prove this we will once again use the representation in terms of complex spherical harmonics, to have a simplified notation for the Wigner $D$-matrices. Analogously to what done for the same case of proper tensors, we align the $\hat{z}$ axis of our frame of reference to the input space $S$. This shows that also for pseudotensors $\Theta_\lambda$, only the $\mu = 0$ component does not vanish, namely, they must possess cylindrical symmetry. We now apply either an inversion around the $xy$-plane or a rotation of $\pi$ around the $y$-axis. The two operations bring the input in the same configuration, as they both cause the inversion $r_i \to -r_i$ for all $i$. However, exploiting equivariance and the explicit expression for inversions, on the one hand we obtain

$$\Theta_\lambda(\{-r_i\}) = -(-1)^\lambda \Theta_\lambda(\{-r_i\}). \tag{D.3}$$

On the other hand, the Wigner $D$-matrix that realizes the rotation is $D^\lambda_{mm'}(0, \pi, 0) = \delta_{mm'}(-1)^{\lambda-\mu}$. Since we have only the $\mu = 0$ component, for the case of a rotation (and exploiting equivariance once more) we obtain

$$\Theta_\lambda(\{-r_i\}) = (-1)^\lambda \Theta_\lambda(\{-r_i\}). \tag{D.4}$$

Comparing the last two equations, we conclude that the only possibility is $\Theta_\lambda = 0$. Finally, we can notice that all the vectors in the input space are aligned and thus all the terms $(r_i \widetilde{\otimes} r_j)_1$ vanish. Therefore, our formulation is compatible with identically vanishing pseudotensors for the $\dim S = 1$ case, concluding the proof of Proposition 4. The proof of Corollary 5 is the same as the one used for Corollary 2. □

**Proof of Proposition 6.** The following proof heavily relies on the results of Proposition (3) and its own proof. Indeed, the main idea is to use the Cartesian (padded) tensor $B^{q_0 q_1 \ldots q_\lambda}$ and follow the contraction path indicated in Eq. (C.4), but for the last coupling, without contracting the component that transform as a pseudotensor. The next step is analogous to the introduction of the arbitrary vector $\hat{s}$. However, because we have to contract by a pseudotensor, we will use two unit vectors, $\hat{a}$ and $\hat{b}$, and contract the free component with the pseudovector $(\hat{a} \widetilde{\otimes} \hat{b})_1$. Explicitly, we take

$$T_{(\lambda-1)\mu'}(r_1, \ldots r_n, \hat{a}, \hat{b}) := \sum_q (\hat{a} \widetilde{\otimes} \hat{b})_{1q} \left[ \sum_{q_0 q_1} C^{1q}_{1q_0 1q_1} \left( \sum_{q_2 \ldots q_\lambda} \mathcal{C}^{(\lambda-1)\mu'}_{q_2 \ldots q_\lambda} B^{q_0 q_1 \ldots q_\lambda} \right) \right].$$

This object has three core properties: on the one hand not identically vanish as the expression in square brackets is on the coupling path that leads to $i\Theta_\lambda$ in Eq. (C.3). on the other hand, it behaves as a proper harmonic tensor of angular momentum $(\lambda - 1)$, as it can be seen from the facts that the only free indices are obtained by a maximal coupling (in the round brackets), and that the pseudotensor components are contracted with a pseudotensor. Finally, since $T_{(\lambda-1)}$ is a linear transformation of $B^{q_0 q_1 \ldots q_\lambda}$, which in turn is obtained by a linear transformation of $\Theta_\lambda$, is polynomial in the input.

Also, using again the relation between unit vectors and spherical harmonics, we have the following integral

$$\int_{\Omega^2} d\hat{a}\, d\hat{b}\, (\hat{a} \widetilde{\otimes} \hat{b})_{1q'} (Y_1(\hat{a}) \widetilde{\otimes} Y_1(\hat{b}))_{1q} = \frac{4\pi}{3} \delta_{qq'}, \tag{D.5}$$

with $(Y_1(\hat{a}) \widetilde{\otimes} Y_1(\hat{b}))_{1q} = \sum_{q_1 q_2} C^{1q}_{1q_1 1q_2} Y_1^{q_1}(\hat{a}) Y_1^{q_2}(\hat{b})$, and where each of the two integration variables runs on the surface of the sphere. Crucially, this integral can be used to select the $q$-th component from the tensor $T_{\lambda-1}$. Indeed, we can recover the components of the component of the pseudotensor by means of the following chain of operations

$$\Theta_{\lambda\mu} = -i\frac{3}{4\pi} \sum_{\mu' q} C^{\lambda\mu}_{(\lambda-1)\mu' 1q} \int_{\Omega^2} d\hat{a}\, d\hat{b}\, T_{(\lambda-1)\mu'}(r_1, \ldots r_n, \hat{a}, \hat{b}) \left(Y_1(\hat{a}) \widetilde{\otimes} Y_1(\hat{b})\right)_{1q}, \tag{D.6}$$

where the integral selects the correct component which is then contracted by means of a CG coefficient into the maximal coupling.

However, $T_{(\lambda-1)\mu'}$ is a proper tensor and polynomial in its inputs. In particular, it is homogeneous of degree 1 in both $\hat{a}$

and $\hat{\boldsymbol{b}}$. Therefore we can apply Proposition 4 and write

$$\begin{aligned}
\boldsymbol{T}_{(\lambda-1)}(\boldsymbol{r}_1,\dots\boldsymbol{r}_n,\hat{\boldsymbol{a}},\hat{\boldsymbol{b}}) = &\sum_{i_1\dots i_{\lambda-1}} h^{(0)}_{i_1\dots i_{\lambda-1}}(\boldsymbol{r}_1,\dots,\boldsymbol{r}_n,\hat{\boldsymbol{a}},\hat{\boldsymbol{b}})\big(\boldsymbol{r}_{i_1}\widetilde{\otimes}\dots\widetilde{\otimes}\boldsymbol{r}_{i_{\lambda-1}}\big)_{\lambda-1} \\
&+ \sum_{i_1\dots i_{\lambda-2}} h^{(1)}_{i_1\dots i_{\lambda-2}}(\boldsymbol{r}_1,\dots,\boldsymbol{r}_n,\hat{\boldsymbol{a}})\big(\hat{\boldsymbol{b}}\widetilde{\otimes}\hat{\boldsymbol{r}}_{i_1}\widetilde{\otimes}\dots\widetilde{\otimes}\hat{\boldsymbol{r}}_{i_{\lambda-2}}\big)_{\lambda-1} \\
&+ \sum_{i_1\dots i_{\lambda-2}} h^{(2)}_{i_1\dots i_{\lambda-2}}(\boldsymbol{r}_1,\dots,\boldsymbol{r}_n,\hat{\boldsymbol{b}})\big(\hat{\boldsymbol{a}}\widetilde{\otimes}\hat{\boldsymbol{r}}_{i_1}\widetilde{\otimes}\dots\widetilde{\otimes}\hat{\boldsymbol{r}}_{i_{\lambda-2}}\big)_{\lambda-1} \\
&+ \sum_{i_1\dots i_{\lambda-3}} h^{(3)}_{i_1\dots i_{\lambda-3}}(\boldsymbol{r}_1,\dots,\boldsymbol{r}_n)\big(\hat{\boldsymbol{a}}\widetilde{\otimes}\hat{\boldsymbol{b}}\widetilde{\otimes}\hat{\boldsymbol{r}}_{i_1}\widetilde{\otimes}\dots\widetilde{\otimes}\hat{\boldsymbol{r}}_{i_{\lambda-3}}\big)_{\lambda-1},
\end{aligned} \quad \text{(D.7)}$$

for some polynomial functions $h^{(0)}$, $h^{(1)}$, $h^{(2)}$, and $h^{(3)}$. Note that this argument holds also for $\lambda = 1$, where only the first function $h^{(0)}$ is retained.

Now the proof is obtained by retracing of the same points of the one for Proposition 4, with the only difference that everything is done on two vectors at the time. For example, exploiting the fact that $\boldsymbol{T}_{(\lambda-1)}$ is homogeneous of degree 1 in $\hat{\boldsymbol{a}}$ and $\hat{\boldsymbol{b}}$, and evoking the fundamental theorem of invariant theory, we can write

$$h^{(0)}_{i_1\dots i_{\lambda-1}}(\boldsymbol{r}_1,\dots,\boldsymbol{r}_n,\hat{\boldsymbol{a}},\hat{\boldsymbol{b}}) = \sum_{i_\lambda i_{\lambda+1}} f_{i_1\dots i_\lambda i_{\lambda+1}}(\boldsymbol{r}_1,\dots,\boldsymbol{r}_n)(\hat{\boldsymbol{a}}\cdot\boldsymbol{r}_{i_\lambda})(\hat{\boldsymbol{b}}\cdot\boldsymbol{r}_{i_{\lambda+1}}), \quad \text{(D.8)}$$

with $f_{i_1\dots i_\lambda i_{\lambda+1}}$ polynomial functions. Performing the integral of Eq. (D.6) now has two effects: On the one hand, it removes the dependence on the unit vectors $\hat{\boldsymbol{a}}$ and $\hat{\boldsymbol{b}}$. On the other hand, in place of the scalar products we obtain the pseudovector $(\boldsymbol{r}_{i_\lambda}\widetilde{\otimes}\boldsymbol{r}_{i_{\lambda+1}})_1$. Thus, the left-most CG contraction of Eq. (D.6), being on the path of maximal coupling, leads to the emergence of $\big((\boldsymbol{r}_{i_\lambda}\widetilde{\otimes}\boldsymbol{r}_{i_{\lambda+1}})_1\widetilde{\otimes}\dots\widetilde{\otimes}\boldsymbol{r}_{i_{\lambda-1}}\big)_\lambda$. In complete analogy to what discussed in the proof for Proposition 4, this is the only term that can be retained: All the others, after the combined application of the fundamental theorem, the action of the integral and the CG contraction, give raise to terms that do not contain enough vector to be equivariantly contracted into a $\lambda$-channel. Thus, dropping everything else and redefining the functions $f$ to match the definition of the statement, we obtain

$$\boldsymbol{\Theta}_\lambda = \mathrm{i}\sum_{i_0\dots i_\lambda} f_{i_0\dots i_\lambda}(\{\boldsymbol{r}_i\})\big((\boldsymbol{r}_{i_0}\widetilde{\otimes}\boldsymbol{r}_{i_1})_1\widetilde{\otimes}\dots\widetilde{\otimes}\boldsymbol{r}_{i_\lambda}\big)_\lambda,$$

as per statement. □

## E  Expansion over three vectors of proper tensors

In this section we derive the expressions of Eqs. (21)-(23) in the main text, where it is assumed that we can use just three vectors $\boldsymbol{q}_1$, $\boldsymbol{q}_2$ and $\boldsymbol{q}_3$ to generate the whole input space $S$. This is analogous to investigating what would happen by considering a minimal basis or the minimal set of vectors that constitute a frame of reference. The cases in which this approximation fails are discussed in the main text.

We are defining an expansion in terms of three vectors $\boldsymbol{q}_1$, $\boldsymbol{q}_2$ and $\boldsymbol{q}_3$ only, by using Prop. 2:

$$\boldsymbol{T}_\lambda(\{\boldsymbol{q}_\alpha\};\{\boldsymbol{r}_i\}) = \sum_{\substack{l_1,l_2,l_3=0 \\ l_1+l_2+l_3=\lambda}}^{\lambda} \Phi_{l_1 l_2 l_3}(\{\boldsymbol{r}_i\})\big(\boldsymbol{q}_1^{\widetilde{\otimes}l_1}\widetilde{\otimes}\boldsymbol{q}_2^{\widetilde{\otimes}l_2}\widetilde{\otimes}\boldsymbol{q}_3^{\widetilde{\otimes}l_3}\big)_\lambda. \quad \text{(E.1)}$$

If $\dim(S) = 1$ we need only one vector $\boldsymbol{q}_1$. If instead the dimensionality is 2, it is necessary to introduce the second vector $\boldsymbol{q}_2$, which we assume not to be collinear with $\boldsymbol{q}_1$. Finally, if the span is the full space, then we introduce the third vector $\boldsymbol{q}_3$, such as

$$\boldsymbol{q}_3 = c_1\boldsymbol{q}_1 + c_2\boldsymbol{q}_2 + c_{12}(\boldsymbol{q}_1\times\boldsymbol{q}_2), \quad \text{(E.2)}$$

for some real coefficients $c_1$, $c_2$ and $c_{12}$. We remark that the coefficient $c_{12}$ must be a pseudoscalar, however this will not introduce any issue because, as it will be shortly shown, we will always use an even product of $c_{12}$ coefficients. This equation holds in general by allowing $c_{12}$ to be 0 whenever $\dim S = 2$. We will not explicitly enforce these condition, which will be assumed to be learned by the model because of the explicit expression that we will use, and that we are now going to derive. Let us firstly investigate the case for $\lambda \geq 3$, and the cases $\lambda = 1$ and $\lambda = 2$ will be discussed at the end of this section. Plugging Eq. (E.2) into (E.1), and using the very same argument used in the proof of Prop. 1, from contractions of the vector $\boldsymbol{q}_3$ we obtain only three types of terms: The first one is obtained when the resulting CG contraction is written in terms of the vectors $\boldsymbol{q}_1$ or $\boldsymbol{q}_2$ only. This can be trivially reabsorbed in cases not spawning from $\boldsymbol{q}_3$ by means of a re-definition of the function $\Psi_{l_1 l_2 l_3}$. The second case is the one in which $\boldsymbol{q}_1\times\boldsymbol{q}_2$ appears an even number of times. Recalling that the spherical representation of the cross product is given by $\mathrm{i}(\boldsymbol{q}_1\widetilde{\otimes}\boldsymbol{q}_2)_1$, we have that this case includes, in the construction of the full CG contraction, terms like

$$i^{2s}(q_1 \widetilde{\otimes} q_2)_1^{\widetilde{\otimes} 2s} = (-1)^s \big((q_1 \widetilde{\otimes} q_2)_1 \widetilde{\otimes} (q_1 \widetilde{\otimes} q_2)_1\big)_2^{\widetilde{\otimes} s}.$$

Using now the first fundamental decomposition of Eq. (7) in the main text, we recognize that also this term can be written as CG contractions of $q_1$ and $q_2$, with all the scalar terms contributing in further re-definitions of the functions $\Psi_{l_1 l_2 l_3}$. Finally, the last case is when the $q_1 \times q_2$ appears an odd number of times, where we can follow again the same rationale of the even case, but, this time, we can observe that one cross product cannot be recombined as a linear combination of $q_1$ and $q_2$. For this remaining cross product, we can reintroduce $q_3$ by means of Eq. (E.2) and go back to an expression in term of $q_1$, $q_2$ and $q_3$ (this operation also reabsorbs all the remaining pseudoscalars $c_{12}$). Factorizing again all the contributions depending on $q_1$ and $q_2$ we are then led to an expression like

$$T_\lambda(\{q_\alpha\}; \{r_i\}) = \sum_{l=0}^{\lambda} f_l(\{r_i\}) \big(q_1^{\widetilde{\otimes} l} \widetilde{\otimes} q_2^{\widetilde{\otimes}(\lambda-l)}\big)_\lambda + \sum_{l=0}^{\lambda-1} g_l(\{r_i\}) \big(q_1^{\widetilde{\otimes} l} \widetilde{\otimes} q_2^{\widetilde{\otimes}(\lambda-l-1)} \widetilde{\otimes} q_3\big)_\lambda, \qquad \text{for } \lambda \geq 3. \tag{E.3}$$

It is important to notice that the above expression requires a set of $\lambda+1$ scalar functions $f_l(\{r_i\})$, and a set of $\lambda$ scalar functions $g_l(\{r_i\})$, hence we obtained an harmonic tensor of angular momentum $\lambda$ (possessing $2\lambda+1$ components) by means of $2\lambda+1$ scalar functions. Before proceeding to the cases for $\lambda = 1, 2$, let us briefly discuss what happens when $\dim(S) = 2$ or $\dim(S) = 1$. For the former case, we have that $q_3$ can be written as a linear combination of the other two vectors. This implies that the second term of the expression above can be re-absorbed into the first one, leading to only $\lambda+1$ truly independent scalar functions. Analogously, if $d = 1$, then there is only one independent vector, leading to only one independent scalar function. This shows that one can infer the number of required independent functions $\{f_l\}$ and $\{g_l\}$ by investigating the linear independence of the vectors $\{q_\alpha\}$. However, unless the set of input vectors $\{r_i\}$ has some known and fixed constraint, it is better to adopt the most general expression available.

The cases for $\lambda = 2$ and $\lambda = 1$ must be treated differently as they allow to treat only two vectors of one at the time, meaning that the division in even and odd terms cannot be carried out.

• **Case for $\lambda = 2$.** Directly from Eq.(E.1), we obtain

$$\begin{aligned}T_2(\{q_\alpha\}; \{r_i\}) =\;& \Psi_{200}(\{r_i\})(q_1 \widetilde{\otimes} q_1)_2 + \Psi_{020}(\{r_i\})(q_2 \widetilde{\otimes} q_2)_2 + \Psi_{002}(\{r_i\})(q_3 \widetilde{\otimes} q_3)_2 \\ &+ \Psi_{110}(\{r_i\})(q_1 \widetilde{\otimes} q_2)_2 + \Psi_{101}(\{r_i\})(q_1 \widetilde{\otimes} q_3)_2 + \Psi_{011}(\{r_i\})(q_2 \widetilde{\otimes} q_3)_2.\end{aligned} \tag{E.4}$$

We now notice that the term $(q_3 \widetilde{\otimes} q_3)_2$ can be re-absorbed in the other five by applying Eq. (E.2) and the identity of Eq. (7) from the main text. In this way, by renaming the scalar functions, we obtain the following general expression

$$T_2(\{q_\alpha\}; \{r_i\}) = \sum_{\alpha=1}^{2} f_\alpha(\{r_i\})(q_\alpha \widetilde{\otimes} q_\alpha)_2 + \sum_{\substack{\alpha_1,\alpha_2=1 \\ \alpha_1<\alpha_2}}^{3} g_{\alpha_1 \alpha_2}(\{r_i\})(q_{\alpha_1} \widetilde{\otimes} q_{\alpha_2})_2, \tag{E.5}$$

which shows that, as expected, we need only 5 scalar functions.

• **Case for $\lambda = 1$.** This simpler case is again obtain from a direct application of Eq. (E.1). By again enumerating the scalar functions with an $\alpha$ index we can write

$$T_1(\{q_\alpha\}; \{r_i\}) = \sum_{\alpha=0}^{2} f_\alpha(\{r_i\}) q_\alpha, \tag{E.6}$$

which is the same result of the one obtained in Ref. 20 for an input of 3 vectors.

## F Expansion over three vectors of pseudo tensors

In this section we will generalize the results for the 3-vector case to the expansion of pseudo-vectors, again separating the cases $\lambda = 1$, $\lambda = 2$ and $\lambda \geq 3$.

• **Case for $\lambda = 1$.** This case is obtained directly from Eq. (18) in the main text and reads

$$\Theta_1(\{q_\alpha\}; \{r_i\}) = i\Big[f_1(\{r_i\})(q_1 \widetilde{\otimes} q_2)_1 + f_2(\{r_i\})(q_1 \widetilde{\otimes} q_3)_1 + f_3(\{r_i\})(q_2 \widetilde{\otimes} q_3)_1\Big], \tag{F.1}$$

with only three scalar functions, as expected from the degrees of freedom of this case. The imaginary unit i is present to extract real components as the CG contractions in the square brackets are purely imaginary.

- **Case for $\lambda = 2$.** From Eq. (18) in the main body, we would have 9 terms to consider. Here we will prove that the linearly independent terms are only five. The general form of these terms is $(q_{\alpha_1} \widetilde{\otimes} q_{\alpha_2})_1 \widetilde{\otimes} q_{\alpha_3})_2$, where the $\alpha_3$ can assume any value in $\{1, 2, 3\}$, while the first two indexes are constrained by $\alpha_1 > \alpha_2$. Let us analyze all these terms: the first two are $((q_1 \widetilde{\otimes} q_2)_1 \widetilde{\otimes} q_1)_2$ and $((q_1 \widetilde{\otimes} q_2)_1 \widetilde{\otimes} q_2)_2$, which are always independent (when $\dim(S) = 1$ these terms are identically zero, in accordance with the fact that no equivariant pseudotensor can be defined). The next term is $((q_1 \widetilde{\otimes} q_2)_1 \widetilde{\otimes} q_3)_2$ which can be written as

$$((q_1 \widetilde{\otimes} q_2)_1 \widetilde{\otimes} q_3)_2 = c_1((q_1 \widetilde{\otimes} q_2)_1 \widetilde{\otimes} q_1)_2 + c_2((q_1 \widetilde{\otimes} q_2)_1 \widetilde{\otimes} q_2)_2 + i\sqrt{2}c_{12}((q_1 \widetilde{\otimes} q_2)_1 \widetilde{\otimes} (q_1 \widetilde{\otimes} q_2)_1)_2, \tag{F.2}$$

by means of Eq. (E.2) and where we used $(q_1 \times q_2) = i\sqrt{2}(q_1 \widetilde{\otimes} q_2)_1$. The first to addends of the right-hand side have already be taken into account, therefore the only really independent component is $((q_1 \widetilde{\otimes} q_2)_1 \widetilde{\otimes} (q_1 \widetilde{\otimes} q_2)_1)_2$, where $c_{12}$ is a pseudoscalar. Now we can use Eq. (A.1) and get

$$((q_1 \widetilde{\otimes} q_2)_1 \widetilde{\otimes} (q_1 \widetilde{\otimes} q_2)_1)_2 = \frac{1}{2}\Big[|q_1|^2(q_2 \widetilde{\otimes} q_2)_2 + |q_2|^2(q_1 \widetilde{\otimes} q_1)_2 - 2(q_1 \cdot q_2)(q_1 \widetilde{\otimes} q_2)_2\Big], \tag{F.3}$$

which suggests that the remaining independent elements are $(q_1 \widetilde{\otimes} q_1)_2$, $(q_2 \widetilde{\otimes} q_2)_2$ and $(q_1 \widetilde{\otimes} q_2)_2$. However, this would also indicate that the dependence on the vector $q_3$ is provided only by the pseudoscalar $c_{12}$. This is indeed the case, as can be seen from Eq. (E.2)

$$c_{12} = \frac{(q_1 \times q_2) \cdot q_3}{\|q_1 \times q_2\|^2} = -i\sqrt{\frac{3}{2}} \frac{((q_1 \widetilde{\otimes} q_2)_1 \widetilde{\otimes} q_3)_0}{\|(q_1 \widetilde{\otimes} q_2)_1\|^2}, \tag{F.4}$$

where we defined all the quantities in terms of the CG contractions only. We also remark that $(q_1 \times q_2) \cdot q_3 = \det\{q_1, q_2, q_3\}$ is the determinant of the matrix formed by the three vectors, and that we define $c_{12} = 0$ every time $(q_1 \widetilde{\otimes} q_2)_1 = 0$. Since this behavior is fully encoded in the determinant alone, we will keep track only of the pseudoscalar $((q_1 \widetilde{\otimes} q_2)_1 \widetilde{\otimes} q_3)_0$.

We can now move to the remaining 6 terms of the form $((q_\alpha \widetilde{\otimes} q_3)_1 \widetilde{\otimes} q_\beta)_2$ with $\alpha \in \{1, 2\}$ and $\beta \in \{1, 2, 3\}$. From Eq. (E.2), and considering the cases $\alpha = 1, 2$ simultaneously, we obtain

$$(q_{1/2} \widetilde{\otimes} q_3)_1 = \pm c_{2/1}(q_1 \widetilde{\otimes} q_2)_1 - \frac{1}{2} c_{12}\Big[(q_{1/2} \cdot q_2)q_1 - (q_1 \cdot q_{1/2})q_2\Big], \tag{F.5}$$

where we used the well-known identity for the triple product $a \times (b \times c) = (a \cdot c)b - (a \cdot b)c$. Coupling this again with $q_1$ and $q_2$ does bring terms that have been already considered in the previous expansions. The situation varies slightly when coupling with $q_3$ but we can use again Eq. (E.2) to relegate the dependence with respect to $q_3$ to the pseudoscalar. Therefore, we just showed that the five linearly independent terms are divided in two groups, the first containing

$$((q_1 \widetilde{\otimes} q_2)_1 \widetilde{\otimes} q_1)_2, \quad \text{and} \quad ((q_1 \widetilde{\otimes} q_2)_1 \widetilde{\otimes} q_2)_2 \tag{F.6}$$

where only pseudovectors appear. These terms are non-zero even when $\dim(S) = 2$. The second group containing

$$((q_1 \widetilde{\otimes} q_2)_1 \widetilde{\otimes} q_3)_0 (q_1 \widetilde{\otimes} q_1)_2, \quad ((q_1 \widetilde{\otimes} q_2)_1 \widetilde{\otimes} q_3)_0 (q_2 \widetilde{\otimes} q_2)_2, \quad ((q_1 \widetilde{\otimes} q_2)_1 \widetilde{\otimes} q_3)_0 (q_1 \widetilde{\otimes} q_2)_2, \tag{F.7}$$

which encode the pseudoscalar and thus go to zero unless the input space spans the whole 3d space. Therefore, the general formula for the pseudotensor becomes

$$\Theta_2(\{q_\alpha\}; \{r_i\}) = i\Bigg[\sum_{\alpha=1}^{2} f_\alpha(\{r_i\})((q_1 \widetilde{\otimes} q_2)_1 \widetilde{\otimes} q_\alpha)_2 \\ + ((q_1 \widetilde{\otimes} q_2)_1 \widetilde{\otimes} q_3)_0 \Big(g_1\{r_i\})(q_1 \widetilde{\otimes} q_1)_2 + g_2(\{r_i\})(q_2 \widetilde{\otimes} q_2)_2 + g_3(\{r_i\})(q_1 \widetilde{\otimes} q_2)_2\Big)\Bigg], \tag{F.8}$$

for some $f_\alpha(\{r_i\})$ and $g_\beta(\{r_i\})$ (proper) scalar functions. We remark that only 5 scalar functions are required, as expected from the dimensionality of this $\lambda = 2$ space. Moreover, this shows that if $\dim(S) = 2$ the degrees of freedom reduce to only two since the pseudoscalar $((q_1 \widetilde{\otimes} q_2)_1 \widetilde{\otimes} q_3)_0$ goes to 0.

- **Case for $\lambda \geq 3$.** This general case can be seen as an extension of the $\lambda = 2$ case and the analogous derivation for proper tensors. Indeed, the general result of Eq. (18) in the main text reads

$$\Theta_\lambda(\{q_\alpha\}; \{r_i\}) = i \sum_{i_0 > i_1} \sum_{i_2 \geq \ldots \geq i_\lambda} f_{i_0 \ldots i_\lambda}(\{r_i\}) \big((q_{i_0} \widetilde{\otimes} q_{i_1})_1 \widetilde{\otimes} q_{i_2} \ldots \widetilde{\otimes} q_{i_\lambda}\big)_\lambda, \tag{F.9}$$

with all the indexes in $\{1, 2, 3\}$. From this we can read the terms $(q_{i_2} \widetilde{\otimes} \ldots \widetilde{\otimes} q_{i_\lambda})_{\lambda-1}$ which are proper harmonic $(\lambda-1)$-tensors. Therefore, using the analogous results for proper tensors (see Eqs. (E.5) and Eqs. (E.3)) we have that the

vector $q_3$ can appear at most in the proper tensorial part. Because the case $\lambda = 2$ proves that each time the tensor $q_3$ appears (in the pseudovector, in the proper tensor part or in both) the expression can be casted in terms of $q_1$ and $q_2$ only but with the presence of the pseudoscalar, the total expression gets partitioned in two parts: one containing the pseudovector ($q_1 \widetilde{\otimes} q_2$) and one containing only proper tensors multiplied by the pseudoscalar. Explicitly, the expansion for the pseudotensor becomes

$$\Theta_\lambda(\{q_\alpha\}; \{r_i\}) = i\left[\sum_{l=0}^{\lambda-1} f_l(\{r_i\})\big((q_1 \widetilde{\otimes} q_2)_1 \widetilde{\otimes} q_1^{\widetilde{\otimes}(\lambda-l-1)} \widetilde{\otimes} q_2^{\widetilde{\otimes} l}\big)_\lambda + \big((q_1 \widetilde{\otimes} q_2)_1 \widetilde{\otimes} q_3\big)_0 \sum_{l=0}^\lambda g_l(\{r_i\})\big(q_1^{\widetilde{\otimes}(\lambda-l)} \widetilde{\otimes} q_2^{\widetilde{\otimes} l}\big)_\lambda\right],$$
(F.10)

for some $f_l(\{r_i\})$ and $g_l(\{r_i\})$ (proper) scalar functions. We can again appreciate how the number of scalar functions is $(2\lambda + 1)$, and that we have a partition in terms that contain the pseudoscalar (which goes to zero when the input is coplanar) and in terms that contain only the pseudovector. Also, the vector $q_3$ does not appear in any CG contraction but in the pseudoscalar. Finally, comparing this with the analogous result for proper tensors of Eq. (E.3), we can notice how the structure is quite similar, with the role of the vector $q_3$ replaced by the pseudovector $(q_1 \widetilde{\otimes} q_2)_1$, and the cases not dependent on $q_3$ multiplied by the pseudoscalar in order to keep the correct behavior under inversion.

## G On the permutational invariance

In this section we will derive Eq. (20) in the main text, for permutationally invariant tensors with $\lambda = 3$. We will focus on this special case since the procedure outlined here can be trivially extended to the general case by following the very same steps outlined here. We will not report the general expression since it is cumbersome and of difficult reading, without providing any significant additional information. The following derivation is almost identical to the one in Ref. 20, but applied to maximal CG contraction. The starting point is the result of Eq. (16) from Coroll. 2, where we partition the expression with respect to how many times the same positions appear in the maximal coupling:

$$\begin{aligned}T_3(\{r_i\}) &= \sum_{i_1 \leq i_2 \leq i_3} f_{i_1 i_2 i_3}(\{r_i\})(r_{i_1} \widetilde{\otimes} r_{i_2} \widetilde{\otimes} r_{i_3})_3 \\ &= \sum_{i_1} f_{i_1 i_1 i_1}(\{r_i\}) r_{i_1}^{\widetilde{\otimes} 3} + \sum_{\substack{i_1 i_2 \\ i_1 \neq i_2}} f_{i_1 i_1 i_2}(\{r_i\})(r_{i_1}^{\widetilde{\otimes} 2} \widetilde{\otimes} r_{i_2})_3 + \sum_{i_1 < i_2 < i_3} f_{i_1 i_2 i_3}(\{r_i\})(r_{i_1} \widetilde{\otimes} r_{i_2} \widetilde{\otimes} r_{i_3})_3,\end{aligned}$$ (G.1)

where the second term has been obtained by exploiting the commutativity of the maximal coupling and the symmetry of the scalar functions under indices permutation. Indicating with $S_n$ the set of all possible permutation of $n$-elements, and performing a permutational average, we have

$$T_3(\{r_i\}) = \frac{1}{n!} \sum_{\sigma \in S_n} T_3(r_{\sigma(1)}, \ldots, r_{\sigma(n)}) = \frac{1}{n!} \sum_{\sigma \in S_n} T_3(r_\sigma),$$ (G.2)

where the last equality defines the shorthand $(r_\sigma) := (r_{\sigma(1)}, \ldots, r_{\sigma(n)})$ for the permuted inputs. We can now separate the average over all the three addends. In particular, the average of the first one is

$$\frac{1}{n!} \sum_{\sigma \in S_n} \sum_j f_j^{(0)}(r_\sigma) r_{\sigma(j)}^{\widetilde{\otimes} 3} = \frac{1}{n!} \sum_j \sum_k \sum_{\sigma \in \Pi_{j \to k}} f_j^{(0)}(r_\sigma) r_k^{\widetilde{\otimes} 3} = \sum_k \left(\frac{1}{n!} \sum_j \sum_{\sigma \in \Pi_{j \to k}} f_j^{(0)}(r_\sigma)\right) r_k^{\widetilde{\otimes} 3},$$ (G.3)

where we defined the $f_j^{(0)} := f_{jjj}$ for readability and we introduced the set of all permutation that map $j$ in $k$, $\Pi_{j \to k} := \{\sigma \in S_n : \sigma(j) = k\}$. The steps above are so far identical to the analogous ones of Ref. 20 and leads to the same conclusions: Firstly, the term in the curly brackets is symmetric under any permutation $\tau_k$ of indices that keep the $k$-th fixed (the stabilizer of the $k$-th index, $\tau_k(k) = k$). This can be easily seen by noticing that, if $\sigma \in \Pi_{j \to k}$, then also the composition $(\tau_k \circ \sigma) \in \Pi_{j \to k}$. Then one can define the function

$$\widetilde{f}_k^{(0)}(r_k, [r_i]_{i \neq k}) := \frac{1}{n!} \sum_j \sum_{\sigma \in \Pi_{j \to k}} f_j^{(0)}(r_\sigma),$$ (G.4)

with the convention that the square brackets indicate a permutationally invariant set. Secondly, the function $\widetilde{f}_k^{(0)}$ cannot depend on the index $k$. This can be seen by either performing a further permutational average or, equivalently, by noticing that having index-dependent functions would allow to label the position of the inputs of $T_3(\{r_i\})$, because the tensors $r_k^{\widetilde{\otimes} 3}$ would be multiplied by different scalar functions. Thus one can further define $\widetilde{f}_k^{(0)}(r_k, [r_i]_{i \neq k}) =: f_0(r_k, [r_i]_{i \neq k})$, which directly leads to the first addend of Eq. (20) in the main text. The very same argument can be used on the second addend since it holds that, symbolically,

$$\sum_{\sigma \in S_n} = \sum_{j_1} \sum_{\sigma \in \Pi_{i_1 \to j_1}} = \sum_{\substack{j_1 j_2 \\ j_1 \neq j_2}} \sum_{\sigma \in \Pi_{(i_1 i_2) \to (j_1 j_2)}},$$ (G.5)

where $\Pi_{(i_1 i_2) \to (j_1 j_2)}$ is the set of all permutations that send the pair of indices $(i_1 i_2)$ into $(j_1 j_2)$. Again, since the double stabilizer $\tau_{j_1 j_2}$ is such that $(\tau_{j_1 j_2} \circ \sigma) \in \Pi_{(i_1 i_2) \to (j_1 j_2)}$, then the function

$$\widetilde{f}^{(1)}_{j_1 j_2}(\boldsymbol{r}_{j_1}, \boldsymbol{r}_{j_2}, [\boldsymbol{r}_i]_{i \notin \{j_1, j_2\}}) := \frac{1}{n!} \sum_{\substack{i_1, i_2 \\ i_1 \neq i_2}} \sum_{\sigma \in \Pi_{(i_1 i_2) \to (j_1 j_2)}} f_{i_1 i_1 i_2}(\boldsymbol{r}_\sigma), \tag{G.6}$$

is permutationally invariant under the swap of any indices but the $j_1$-th and $j_2$-th ones. Applying the permutational average shows once again that the function $\widetilde{f}^{(1)}_{j_1 j_2}$ cannot depend on the indices $j_1$ and $j_2$, leading to the result

$$\sum_{\substack{i_1 i_2 \\ i_1 \neq i_2}} f_{i_1 i_1 i_2}(\{\boldsymbol{r}_i\})(\boldsymbol{r}_{i_1}^{\widetilde{\otimes} 2} \widetilde{\otimes} \boldsymbol{r}_{i_2})_3 = \sum_{\substack{i_1 i_2 \\ i_1 \neq i_2}} f^{(1)}(\boldsymbol{r}_{i_1}, \boldsymbol{r}_{i_2}, [\boldsymbol{r}_i]_{i \notin \{i_1, i_2\}})(\boldsymbol{r}_{i_1}^{\widetilde{\otimes} 2} \widetilde{\otimes} \boldsymbol{r}_{i_2})_3. \tag{G.7}$$

This can be further simplified by lifting the constraint: The case for $i_1 = i_2$ is again in the form of the first addend of Eq. (20), and thus can be re-absorbed by a simple re-definition of the functions $f_0(\boldsymbol{r}_k, [\boldsymbol{r}_i]_{i \neq k})$.

The last addend is obtained by means of the very same procedure, with the starting point being the relation

$$\sum_{\sigma \in S_n} = \sum_{j_1 j_2 j_3}^{\text{triplets}} \sum_{\sigma \in \Pi_{(i_1 i_2 i_3) \to (j_1 j_2 j_3)}}, \tag{G.8}$$

where the second sum runs over proper triplets of indices. The only difference with the previous cases is that one has also to enforce the permutational invariance with respect to the swap of indices that appear in the maximal coupling (as they are already symmetric). This leads to the full expression of Eq. (20).

Clearly, the approach is trivially generalizable, albeit tedious. In general, the following recipe leads to permutationally invariant expressions for any $\lambda$ of choice: starting from an angular momentum $\lambda$ write all the independent maximal coupling terms with respect to the number of same positions appearing in the coupling. The number of this possible addends is given by the *integer partition* of $\lambda$. Then, average each term on the permutation group. Simplify the expressions by factoring the permutations with respect to their action on the indices appearing in the maximal coupling. In the last two steps, use the stabilizer of these permutations to prove the invariance with respect to all the indices that do not appear in the coupling, and use another average to prove that the functions cannot depend on any of the indices involved. Finally, enforce the remaining symmetries to mirror the one possessed by the CG coupling.

## H  Conventions on the irreducible decomposition of Cartesian tensors

In this short section we discuss the convention followed for the numerical experiments in the main text to decompose Cartesian tensors into their irreducible spherical components. We treat only examples of objects of rank 1, 2 and 3.

- **Rank** 1.   In this case the spherical representation coincides with the one of a vector with rolled indexes. In particular, we follow the convention

$$\begin{cases} y \leftrightarrow -1, \\ z \leftrightarrow 0, \\ x \leftrightarrow 1, \end{cases} \tag{H.1}$$

to move from a Cartesian representation to the spherical representation and back.

- **Rank** 2.   For this case, the decomposition in irreducible components of the general tensor $T$ follows the scheme

$$1 \otimes 1 = 0_+ \oplus 1_- \oplus 2_+. \tag{H.2}$$

Here we kept track also of the behavior under parity, with a subscript $-$ indicating a pseudotensor and a subscript $+$ indicating proper tensor. Since our simulation target only proper tensors, the antisymmetric component for $\lambda = 1$ is absent. In general, we use the CG coefficient to derive the irreducible components of the tensor

$$T^{(\text{irr})}_{\lambda \mu} = \sum_{m_1 m_2} C^{\lambda \mu}_{1 m_1 1 m_2} T^{(\text{roll})}_{m_1 m_2}, \tag{H.3}$$

where $C^{\lambda \mu}_{1 m_1 1 m_2}$ are the CG coefficients for the real representation of spherical harmonics and $T^{(\text{roll})}_{m_1 m_2}$ is the Cartesian tensor where each index has been rolled by means of Eq. (H.1). In this case we provide also the explicit formula for the components. For $\lambda = 0$ is the scalar term and is given by the trace

$$T^{(\text{irr})}_{00} = -\frac{1}{\sqrt{3}}(T_{xx} + T_{yy} + T_{zz}), \tag{H.4}$$

where the left-hand side is the (scalar) spherical representation of the tensor. The components for the $\lambda = 2$ case are given by

$$T^{(\text{irr})}_{2,-2} = \frac{1}{\sqrt{2}}(T_{xy} + T_{yx}),$$

$$T^{(\text{irr})}_{2,-1} = \frac{1}{\sqrt{2}}(T_{yz} + T_{zy}),$$

$$T^{(\text{irr})}_{2,0} = \frac{1}{\sqrt{6}}(2T_{zz} - T_{xx} - T_{yy}), \tag{H.5}$$

$$T^{(\text{irr})}_{2,1} = \frac{1}{\sqrt{2}}(T_{xz} + T_{zx}),$$

$$T^{(\text{irr})}_{2,2} = \frac{1}{\sqrt{2}}(T_{xx} - T_{yy}).$$

- **Rank** 3. For this last case we first follow the decomposition provided by

$$((1 \otimes 1) \otimes 1) = ((0_+ \oplus 1_- \oplus 2_+) \otimes 1) = 1_+ \oplus (0_- \oplus 1_+ \oplus 2_-) \oplus (1_+ \oplus 2_- \oplus 3_+). \tag{H.6}$$

Again, since we target only proper tensors, all the pseudo ones are absent. Moreover, our simulation involved only tensors which are completely symmetric. Therefore among the proper components, the only ones that survive are the maximal one, for $\lambda = 3$, and the linear combination (see Ref. 54, 5)

$$\boldsymbol{S}^{(\text{irr})}_1 = \frac{\sqrt{5}}{3}\boldsymbol{T}^{(\text{irr})}_{0,1} + \frac{2}{3}\boldsymbol{T}^{(\text{irr})}_{1,1}. \tag{H.7}$$

Here we used the definition

$$\left(\boldsymbol{T}^{(\text{irr})}_{L,\lambda}\right)_\mu := \sum_{m_1 m_2 m_3 M} C^{LM}_{1m_1 1m_2} C^{\lambda\mu}_{LM l_3 m_3} T^{(\text{roll})}_{m_1 m_2 m_3}, \tag{H.8}$$

to explicitly address the intermediate coupling channel and thus separate the components with respect to the path of coupling.

## I Derivatives of the MCoV model for the CO$_2$ molecule without the $\lambda$ correction term

This section is devoted to discuss the failure of the model without correction for the case of the CO$_2$ molecule. As shown in Figure 4 of the main text, the spherical expansion centered on the C atom, $\boldsymbol{\rho}^{(C)}_{O,n1}$, goes identically to zero. With the two O atoms aligned on the $z$ axis and at distance $d_1$ and $d_2$ from the $C$ atom, the spherical expansion is

$$\boldsymbol{\rho}^{(C)}_{O,n1} = R_{nl}(d_1)Y_1(\hat{\boldsymbol{z}}) + R_{nl}(d_1)Y_1(-\hat{\boldsymbol{z}}) = \left(R_{nl}(d_1) - R_{n1}(d_2)\right)Y_1(\hat{\boldsymbol{z}}), \tag{I.1}$$

where we used the inversion property of the spherical harmonics, $Y_l(-\hat{\boldsymbol{r}}) = (-1)^l Y_l(\hat{\boldsymbol{r}})$. The spherical expansion manifestly vanish whenever $d_1 = d_2$, as expected from symmetry arguments. Here we are going to investigate the behavior of the model (without correction) under differentiation with respect to one of the distances. This will provide information on the behavior of the model in a neighborhood of the degenerate cases arising whenever $d_1 = d_2$. Since the model is obtained by product of a SOAP-BPNN for the scalar functions and a maximal coupling of the $\{\boldsymbol{q}_\alpha\}_{\alpha=1}^3$ vectors for the equivariant part, the $t$-th derivative with respect to $d_1$ is obtained by linear combinations of terms like

$$\frac{\partial^t}{\partial d_1^t}\left(f(\{\boldsymbol{\rho}_{nl}\})(\boldsymbol{q}_{\alpha_1} \tilde{\otimes} \ldots \tilde{\otimes} \boldsymbol{q}_{\alpha_\lambda})_\lambda\right) = \sum_{k=0}^t \binom{t}{k}\frac{\partial^{t-k} f(\{\boldsymbol{\rho}_{nl}\})}{\partial d_1^{t-k}} \frac{\partial^k (\boldsymbol{q}_{\alpha_1} \tilde{\otimes} \ldots \tilde{\otimes} \boldsymbol{q}_{\alpha_\lambda})_\lambda}{\partial d_1^k} \tag{I.2}$$

where the function $f(\{\boldsymbol{\rho}_{nl}\})$ represents the SOAP-BPNN and with $\alpha_i \in 1,2,3$ for every $i \in 1,\ldots,\lambda$. We implied the dependence of the spherical expansion from the atomic species C and O for readability. Investigating the derivative of the maximal CG coupling we notice that the vectors $\{\boldsymbol{q}_\alpha\}_{\alpha=1}^3$ are obtained by linear combinations of the spherical expansions $\{\boldsymbol{\rho}_{n1}\}$. Thus, it does suffice to differentiate the CG contractions of spherical expansions directly. Explicitly we have

$$\frac{\partial^k}{\partial d_1^k}\left[(\boldsymbol{\rho}_{n_1 1} \tilde{\otimes} \ldots \tilde{\otimes} \boldsymbol{\rho}_{n_\lambda 1})\right] = \sum_{\substack{k_1,\ldots,k_\lambda=0 \\ k_1+\ldots+k_\lambda=k}}^k \binom{k}{k_1,\ldots,k_\lambda}\left[\left(\frac{\partial^{k_1}\boldsymbol{\rho}_{n_1 1}}{\partial d_1^{k_1}} \tilde{\otimes} \ldots \tilde{\otimes} \frac{\partial^{k_\lambda}\boldsymbol{\rho}_{n_\lambda 1}}{\partial d_\lambda^{k_\lambda}}\right)_\lambda\right] \tag{I.3}$$

where we used the generalized Leibniz rule with the multinomial coefficient defined as

$$\binom{k}{k_1,\ldots,k_\lambda} := \frac{k!}{k_1! \cdot \ldots \cdot k_\lambda!} \tag{I.4}$$

From the constraint on the summation we notice that if $k < \lambda$ some of the $k_i$ must be necessarily 0. Therefore, each of the addend in the summation contains at least one spherical expansion which is not differentiated and which goes to zero whenever $d_1 = d_2$, making the whole contribution vanish. Retracing the derivation up to the $t$-th derivative we then deduce that all the derivatives of order up to $\lambda - 1$ are necessarily zero.

This simple analysis, which is trivially generalizable to any highly symmetric configuration, shows that the effect of the degeneracy has a significant effect also on the neighbors of the degeneracy and that this effect is as severe as high is the angular momentum that we are representing.



\end{document}